\documentclass[showpacs,aps,prd,nofootinbib,floatfix,amsmath,amssymb]{revtex4-2}
\usepackage{graphicx}
\usepackage[utf8x]{inputenc}
\usepackage{color}
\usepackage{dsfont}
\usepackage{subfig}
\usepackage{multirow}
\usepackage{hyperref}
\usepackage{slashed}
\usepackage{cancel}%\cancel{} \bcancel{} \xcancel{} \cancelto{value}{expression}
\usepackage{pdfsync}%Inverse search
\usepackage{booktabs}
\usepackage{float}
\begin{document}

\title{Chromomagnetic dipole moment of the top quark in the Bestest Little Higgs model}

\author{J. I. Aranda$^{1}$}
\author{T. Cisneros-P\'erez$^{1}$}
\author{E. Cruz-Albaro$^{1}$}
\author{J. Monta\~no-Dom\'inguez$^{1,2}$}
\author{F. Ram\'irez-Zavaleta$^{1}$}\email{feramirez@umich.mx}
\affiliation{
$^{1}$Facultad de Ciencias F\'isico Matem\'aticas, Universidad Michoacana de San Nicol\'as de Hidalgo,
Av. Francisco J. M\'ugica s/n, C.~P. 58060, Morelia, Michoac\'an, M\'exico.
\\
$^{2}$C\'atedras Conacyt, Av. Insurgentes Sur 1582, Col. Cr\'edito Constructor, Alc. Benito Ju\'arez,
C.~P. 03940, Ciudad de M\'exico, M\'exico.
}

\begin{abstract}
We calculate the chromomagnetic dipole moment of the top quark, $\hat{\mu}_t$, in the context of the Bestest Little Higgs model. This extension of the Standard Model aims to solve the hierarchy problem without fine tuning, by introducing one-loop corrections to the Higgs boson mass through heavy top quark partners and heavy gauge bosons. We found that the largest resulting value for the chromomagnetic is $\hat{\mu}_t\sim 10^{-5}$ and the lowest is around $10^{-6}$, mainly due to the Higgs boson of the Standard Model, which couples to both the top quark of the Standard Model and its heavy partners. Also, we present a wide variety of new Feynman rules involved in our calculation.
\end{abstract}

\maketitle

%%%%%% (1) Motivations $$$$$$
\section{Introduction}
The Little Higgs models \cite{ArkaniHamed:2001nc,ArkaniHamed:2002qy,ArkaniHamed:2002pa,ArkaniHamed:2002qx} offer solutions to the hierarchy problem of the Standard Model (SM), nevertheless, some of these models exhibit issues such as a reduced fine tuning or the violation of the custodial symmetry~\cite{Chang:2003zn,Schmaltz:2008vd}. In this regard, it is know that the Bestest Little Higgs (BLH) model overcomes such difficulties \cite{Schmaltz:2010ac}. In the BLH model occurs a collective symmetry breaking, which involves two breaking scales, $f$ and $F$, where $F>f\sim1$ TeV \cite{Schmaltz:2010ac}, in this way the model solves former dangerous situations, such as the generation of a quartic coupling for the Higgs boson that violates custodial symmetry and the presence of reduced fine-tuning for the top and Higgs sectors. The BLH model gives rise to a new set of degrees of freedom, which are near to the electroweak (EW) scale, promoting a rich phenomenology for the top quark and its heavy BLH partners. In this sense, Ref.~\cite{Godfrey:2012tf} explored the heavy quark sector of the BLH model, where lower mass limits are imposed via CMS experimental results. Additionally, in Ref.~\cite{Kalyniak:2013eva} the BLH parameter space has been constrained from ATLAS and CMS data.

Because of its large mass, the top quark constitutes a bridge between the EW energy scale and the energy scale where the BLH model emerges. Owing to this situation, the top quark interacts extensively with the new degrees of freedom coming from the BLH model~\cite{Schmaltz:2010ac}. Therefore, the search for new physics could be of great interest by considering the physics properties of the top quark, such as its chromoelectromagnetic dipole moments~\cite{Aranda:2018zis}.

In the last two decades, interest about chromomagnetic (CMDM) and chromoelectric (CEDM) dipole moments of the top quark has grown. Firstly, in Ref.~\cite{Davydychev:2000rt} the quantum chromodynamics contribution for the CMDM of the SM quarks, induced at the one-loop level, was partially computed. Subsequently, in Refs.~\cite{Aranda:2018zis,Aranda:2020tox} complete one-loop contributions to the CMDM of the top quark within the SM were calculated. This last SM prediction matches well with the recent experimental measurement achieved by the CMS Collaboration~\cite{Sirunyan:2019eyu,PDG2020}. Specifically, in 2020, the CMS Collaboration, at CERN, published the most recent measurement of the top quark CMDM~ \cite{Sirunyan:2019eyu,PDG2020}
\begin{eqnarray}
\hat{\mu}_t^\mathrm{Exp}=-0.024_{-0.009}^{+0.013}(\mathrm{stat})_{-0.011}^{+0.016}(\mathrm{syst}),\nonumber
\end{eqnarray}
at $95\%$ C.~L., by using $pp$ collisions at the centre-of-mass energy of 13 TeV with an integrated luminosity of 35.9 $fb^{-1}$. In this sense, in Ref. \cite{Aranda:2018zis} a phenomenological analysis of the chromoelectromagnetic properties of the top quark is carried out, by taking into account flavor changing neutral currents in the context of grand unified models. Additionally, the top-quark CMDM has been computed in several extensions of the SM such as the two-Higgs doublet model~\cite{Gaitan:2015aia}, the minimal supersymmetric Standard Model~\cite{Martinez:2001qs,Aboubrahim}, in a few models beyond the SM~\cite{Martinez:2007qf}, in technicolor models~\cite{Appelquist}, in the Littlest Higgs model with T-parity~\cite{Cao:2008qd}, in unparticle physics model~\cite{Martinez:2008hm}, in models with vector-like multiplets~\cite{Ibrahim}, in effective Lagrangians~\cite{Hayreter}, in the two-Higgs doublet model with four fermion generations~\cite{Hernandez-Juarez:2018uow}, and in the reduced 331 model~\cite{Hernandez-Juarez:2020xon}.

From the theoretical point of view, it is known that the CMDM of the SM quarks cannot be well defined in a static scenario, that is to say, when $q^2=0$; it can only be established an off-shell CMDM, which is consistent with an observable given that it is gauge invariant, gauge independent, infrared finite and ultraviolet finite~\cite{Aranda:2020tox,Davydychev:2000rt,Hernandez-Juarez:2020drn,Montano-Dominguez:2021eeg}. Thus, we will present the evaluation of this observable considering an off-shell gluon ($q^2\neq 0$) at large-momentum transfer. In particular, as it happens with the strong running coupling constant $\alpha_s(m_Z^2)$~\cite{Deur:2016tte}, the CMDM of the top quark shall be evaluated just at the energy scale of the $Z$ gauge boson mass. In this paper we calculate the one-loop corrections to the $\bar{t}tg$ vertex in the context of the BLH model. This correction induces the CMDM of the top quark, whereas the respective CEDM is identically zero.

The paper is organized as follows. In Sec. \ref{TF} we briefly discuss the theoretical framework of the BLH model. In Sec. \ref{OLA} we describe the associated amplitudes to the CMDM in the BLH model context. In Sec. \ref{Phenomenology}, we present the numerical results for the CMDM of the top quark in the BLH model. Section \ref{Conclusions} is devoted to our conclusions. Appendix~\ref{Appendix} contains the BLH Feynman rules employed in this work.

%%%%%%%%%%%%%%%%%%%%%%%%%%%%%%
\section{Theoretical framework}\label{TF}

The BLH model is established when a non-linear sigma field $(\Sigma)$ is introduced in the global symmetry $SO(6)_A\times SO(6)_B$~\cite{Schmaltz:2010ac}. At first, the $S0(6)_A\times S0(6)_B$ group is spontaneously broken into the $SO(6)_V$ group, at the energy scale $f$, once $\Sigma$ acquires the vacuum expectation value (VEV)~\cite{Schmaltz:2010ac,Moats:2012laz}
\begin{equation}\label{break6}
 \langle\Sigma\rangle=\textbf{1}_{6\times6}.
\end{equation}
Here, $\langle\Sigma\rangle$ represents a $6\times6$ unit matrix.
Since $SO(6)$ is originated by $15$ symmetry generators, this symmetry breaking mechanism also induces $15$ pseudo-Nambu-Goldstone bosons (PNGBs). The upper $SO(4)$ left-block in each $SO(6)$ can be decomposed into $SU(2)_{LA,B}\times SU(2)_{RA,B}$ symmetry gauge groups~\cite{Schmaltz:2010ac}, respectively.  Afterwards, the diagonal combination of $SU(2)_{LA}$ and $SU(2)_{LB}$ is weakly gauged, being identified as the $SU(2)_L$ gauge group of the SM~\cite{Schmaltz:2010ac}. Additionally, once weakly gauged the diagonal combination of the third component of $SU(2)_{R}$, the SM hypercharge is also found~\cite{Schmaltz:2010ac}. The PNGBs are parameterized around the VEV as follows
\begin{equation}\label{sigma}
 \Sigma=e^{i\Pi/f}e^{2i\Pi_h/f}e^{i\Pi/f},
\end{equation}
where $\Pi$ and $\Pi_h$ are $6\times6$ imaginary anti-symmetric matrices:
\begin{equation}
\Pi=
\begin{pmatrix}
(\phi_aT_L^a+\eta_aT_R^a)_{4\times4} & 0 & 0 \\
0 & 0 & i\sigma/\sqrt{2} \\
0 & -i\sigma/\sqrt{2} & 0
\end{pmatrix},\hspace{0.5cm} (a=1,2,3),
\quad\quad
\Pi_h=\frac{i}{\sqrt{2}}
\begin{pmatrix}
 0_{4\times4} & h_1 & h_2\\
 -h_1^T & 0 & 0\\
 -h_2^T & 0 & 0
\end{pmatrix},
\end{equation}
being
\begin{equation}
 h_j^T=(h_{j1},h_{j2},h_{j3},h_{j4}),  \hspace{0.5cm}(i=1,2).
\end{equation}
The $h_i^T$ are multiplets of the $SO(4)$ group, which can be arranged as complex Higgs doublets~\cite{Schmaltz:2010ac}:
\begin{equation}
 H_i=\frac{1}{\sqrt{2}}
 \begin{pmatrix}
  h_{i3}+ih_{i4}\\
  h_{i1}-ih_{i2}
 \end{pmatrix}.
\end{equation}
Moreover, $\sigma$ is a real singlet field, necessary to achieve a collective quartic coupling. The $T_L^a$ and $T_R^a$ are generators of the $SU(2)_L$ and $SU(2)_R$ gauge groups, respectively, which are within $SO(4)$. The $\phi_a$ field is an electroweak triplet with zero hypercharge. The $\eta_a$ field represents an $SU(2)_R$ triplet, being $\eta_1$ and $\eta_2$ complex singlets with hypercharge and $\eta_3$ a real singlet~\cite{Schmaltz:2010ac}.

In order to keep the custodial symmetry, via a suitable Higgs quartic coupling, it is proposed the following collective quartic potential~\cite{Schmaltz:2010ac}
\begin{equation}\label{p-cuartico}
  V_{quartic}=\frac{1}{4}\lambda_{65}f^4 Tr(P_6\Sigma P_5\Sigma^T)+\frac{1}{4}\lambda_{56}f^4 Tr(P_5\Sigma P_6\Sigma^T)=\frac{1}{4}\lambda_{56}f^4(\Sigma_{56})^2+\frac{1}{4}\lambda_{65}f^4(\Sigma_{65})^2.
\end{equation}
Here, the collective symmetry breaking requires two operators:
\begin{equation}
  P_5=diag(0, 0, 0, 0, 1, 0),\hspace{0.5cm}P_6=diag(0, 0, 0, 0, 0, 1),
\end{equation}
which explicitly break some of the global symmetries, but averting the arising of a Higgs potential~\cite{Schmaltz:2010ac}. The first term in Eq. (\ref{p-cuartico}) promotes the breaking of $SO(6)_A\times SO(6)_B$ towards $SO(5)_{A6}\times SO(5)_{B5}$, where $SO(5)_{An}$ transformations do not act on the nth row or column. In the same way, the second term in Eq. (\ref{p-cuartico}) breaks the global symmetry $SO(6)_A\times SO(6)_B\to SO(5)_{A5}\times SO(5)_{B6}$. Consequently, the two terms in Eq. (\ref{p-cuartico}) break the global symmetry into $SO(4)_A\times SO(4)_B$, where the gauge and custodial symmetries are not explicitly broken~\cite{Schmaltz:2010ac}. Then, by expanding $\Sigma$ as a power series of $1/f$ in Eq. (\ref{p-cuartico}), $V_{quartic}$ can be written as
\begin{equation}\label{pocuartico}
 V_{quartic}=\frac{\lambda_{65}}{2}\left(f\sigma+\frac{1}{\sqrt{2}}h_1^Th_2+\ldots\right)^2+\frac{\lambda_{56}}{2}\left(f\sigma-\frac{1}{\sqrt{2}}h_1^Th_2+\ldots\right)^2.
\end{equation}
From Eq.~(\ref{pocuartico}) a mass term for the scalar $\sigma$ can be extracted
\begin{equation}\label{masa-s}
 m_{\sigma}^2=(\lambda_{65}+\lambda_{56})f^2.
\end{equation}
The apparent presence of a quartic coupling for the Higgses in Eq. (\ref{pocuartico}) can be removed by using the field redefinition $\sigma\rightarrow\sigma\pm h_1^Th_2/\sqrt{2}f$~\cite{Schmaltz:2010ac}. By integrating out $\sigma$, at tree level, in Eq.~(\ref{pocuartico}), it is found the collective quartic potential~\cite{Schmaltz:2010ac}
\begin{equation}\label{lambda0}
V_{quartic}=\frac{\lambda_{56}\lambda_{65}}{\lambda_{65}+\lambda_{56}}
\left(h_1^Th_2\right)^2=\frac{1}{2}\lambda_0\left(h_1^Th_2\right)^2.
\end{equation}

In this regard, the complete scalar potential has the following structure~\cite{Schmaltz:2010ac,Moats:2012laz}
\begin{equation}\label{fullpotential}
 V=V_{quartic}+V_{SB}+V_{B_{\mu}}.
\end{equation}
In order to provide mass for all the scalars, the following symmetry breaking potential, $V_{SB}$, is introduced~\cite{Schmaltz:2010ac,Moats:2012laz}
\begin{equation}\label{vsb}
 V_{SB}=-\frac{f^2}{4}m_4^2Tr\left(\Delta^{\dag} M_{26}\Sigma M^{\dag}_{26}+\Delta M_{26}\Sigma^{\dag} M^{\dag}_{26}\right)-\frac{f^2}{4}\left(m_5^2\Sigma_{55}+m^2_6\Sigma_{66}\right),
\end{equation}
where $\Delta$ symbolizes a field which  breaks the global $SU(2)_C\times SU(2)_D$ symmetry to the diagonal at the energy scale $F$, where $F>f$. As an analogy with the $\Sigma$ field, $\Delta$ is parameterized as
\begin{equation}
 \Delta=e^{2i\Pi_d/F},\hspace{0.5cm}\Pi_d=\chi_a\frac{\tau_a}{2},\hspace{0.5cm}a=1,2,3,
\end{equation}
where $\chi_a$ is a triplet of $SU(2)$. Moreover, $M_{26}$, that is a matrix that connects the $SU(2)$ indices with those of $S0(6)$ of $\Sigma$, is defined as~\cite{Schmaltz:2010ac,Moats:2012laz}
\begin{equation}
 M_{26}=\frac{1}{\sqrt{2}}
 \begin{pmatrix}
  0 & 0 & 1 & i & 0 & 0\\
  1 & -i & 0 & 0 & 0 & 0
 \end{pmatrix}.
\end{equation}
By expanding $\Delta$ in powers of $1/F$ and substituting it in Eq. (\ref{vsb}), it follows that~\cite{Schmaltz:2010ac,Moats:2012laz}
\begin{equation}\label{vsb-masa}
V_{SB}=\frac{1}{2}m_{\phi}^2\phi_a^2
+\frac{1}{2}m_{\eta}^2\eta_a^2+\frac{1}{2}m_1^2h_1^Th_1
+\frac{1}{2}m_2^2h_2^Th_2+\frac{1}{4}(m_5^2+m_6^2)\sigma^2,
\end{equation}
where now the $h_1$, $h_2$, $\phi_a$ and $\eta_a$ fields acquire masses given by
\begin{equation}
 m_{\phi}^2=m_{\eta}^2=m_4^2,\hspace{0.5cm}m_1^2=\frac{1}{2}(m_4^2+m_5^2),
 \hspace{0.5cm}m_2^2=\frac{1}{2}(m_4^2+m_6^2).
\end{equation}

The $V_{B_{\mu}}$ term, in the scalar potential, is added to develop the electroweak symmetry breaking (EWSB)~\cite{Schmaltz:2010ac,Moats:2012laz},
\begin{eqnarray}
 V_{B_{\mu}}&=&m^2_{56}f^2\Sigma_{56}+m^2_{65}f^2\Sigma\nonumber\\
 &=&\sqrt{2}(m^2_{65}-m^2_{56})f\sigma-(m^2_{56}+m^2_{65})h_1^Th_2+\ldots
\end{eqnarray}
Due to $\sigma$ is the heaviest scalar, the full scalar potential, $V$, can be established in the limit $f\gg v$. Then, by minimizing Eq. (\ref{fullpotential}) with respect to $\sigma$ and substituting the solution for $\sigma$ into Eq. (\ref{fullpotential}), this leads to a potential for the Higgs~\cite{Moats:2012laz}
\begin{equation}\label{potencial-h}
 V_{Higgs}=\frac{1}{2}m_1^2h_1^Th_1+\frac{1}{2}m_2^2h_2^Th_2-B_{\mu}h_1^Th_2
 +\frac{\lambda_0}{2}(h_1^Th_2)^2,
\end{equation}
where $ B_{\mu}=2(\lambda_{56}m_{65}^2+\lambda_{65}m_{56}^2)/(\lambda_{56}+\lambda_{65})$ and $\lambda_0=2\lambda_{56}\lambda_{65}/(\lambda_{56}+\lambda_{65})$ (see Eq.~(\ref{lambda0})). To ensure that the potential acquires a minimum, it must be fulfilled that $m_1, m_2 > 0$, whereas EWSB requires that $B_\mu > m_1 m_2$. Thus, these last results illustrate the key idea of the collective symmetry breaking (for more details Ref.~\cite{Moats:2012laz} should be consulted).

In the BLH model, the EWSB arises when the Higgs fields acquire the VEVs:
\begin{eqnarray}
 \langle h_1\rangle^T=(v_1,0,0,0),\\
 \langle h_2\rangle^T=(v_2,0,0,0).
\end{eqnarray}
By considering these VEVs and minimizing the Higgs potential in Eq. (\ref{potencial-h}), it is concluded that
\begin{eqnarray}\label{vevs}
 v_1^2=\frac{1}{\lambda_0}\frac{m_2}{m_1}(B_{\mu}-m_1m_2),\\
 v_2^2=\frac{1}{\lambda_0}\frac{m_1}{m_2}(B_{\mu}-m_1m_2),
\end{eqnarray}
where
\begin{equation}\label{vevsm}
 v^2=v_1^2+v_2^2\simeq(246\;\mathrm{GeV})^2.
\end{equation}
As is usual in little Higgs models, the VEVs can be written in terms of the parameters $v$ and $\tan\beta$~\cite{Schmaltz:2010ac}
\begin{equation}
 \tan\beta=\frac{v_1}{v_2}=\frac{m_2}{m_1}.
\end{equation}

The physical Higgs states include the Goldstone bosons, $G^{\pm}$ and $G^0$, which give mass to SM bosons $W^{\pm}$ and $Z$, respectively, as well as the massive states $H^0$, $A^0$, $H'$, $H^{\pm}$, $\phi^0$, $\phi^{\pm}$, $\eta^0$, $\eta^{\pm}$, and $\sigma$ ~\cite{Schmaltz:2010ac,Kalyniak:2013eva}:
\begin{eqnarray}
 m_{A^0}^2&=&m_{H^{\pm}}^2=m_1^2+m_2^2=\frac{2B_{\mu}}{\sin2\beta}-\lambda_0v^2,\\
 m^2_{H^0,H'}&=&\frac{B_{\mu}}{\sin2\beta}\mp\sqrt{\frac{B^2_{\mu}}{\sin^22\beta}
 -2\lambda_0B_{\mu}v^2\sin2\beta+\lambda_0^2v^4\sin^22\beta},\\
\label{escalar-fi}
m^2_{\phi^0}&=&\frac{16}{3}F^2\frac{3g_A^2g_B^2}{32\pi^2}\ln\left(\frac{\Lambda^2}{m^2_{W'}}\right)+m_4^2\frac{f^4+F^4}{F^2(f^2+F^2)},\\
\label{escalar-fiMm}
m^2_{\phi^{\pm}}&=&\frac{16}{3}F^2\frac{3g_A^2g_B^2}{32\pi^2}\ln\left(\frac{\Lambda^2}{m^2_{W'}}\right)+m_4^2\frac{f^4+f^2F^2+F^4}{F^2(f^2+F^2)},\\
\label{escalar-etaMm}
m^2_{\eta^{\pm}}&=&m_4^2+\frac{3f^2{g'}^2}{64\pi^2}\frac{\Lambda^2}{F^2},\\
m^{2}_{\sigma}&=&(\lambda_{56} + \lambda_{65})f^{2}=2\lambda_0 f^{2} K_\sigma\label{msigma}
\end{eqnarray}

\subsection{Gauge sector}
In the BLH model, the gauge invariant non-linear sigma kinetic terms are given by~\cite{Schmaltz:2010ac,Moats:2012laz}
\begin{equation}\label{lag-norma}
 \mathcal{L}=\frac{f^2}{8}Tr\left(D_{\mu}\Sigma^{\dag}D^{\mu}\Sigma\right)
 +\frac{F^2}{4}Tr\left(D_{\mu}\Delta^{\dag}D^{\mu}\Delta\right),
\end{equation}
where the covariant derivatives are
\begin{eqnarray}
 D_{\mu}\Sigma&=&\partial\Sigma+ig_AA^a_{1\mu}T^a\Sigma-ig_B\Sigma A^a_{2\mu}T^a+ig'\left[B^3_{\mu}T'\,^3,\Sigma\right],\\
 D_{\mu}\Delta&=&\partial_{\mu}\Delta+ig_AA^a_{1\mu}\frac{\tau^a}{2}\Delta-ig_B\Delta A^a_{2\mu}\frac{\tau^a}{2},
\end{eqnarray}
Here, $T^a$ and $T'\,^a$ are the $SO(6)$ generators (see Appendix B in Ref.~\cite{Moats:2012laz}), $g_A$ and $g_B$ are the gauge couplings from $SU(2)_{LA,LB}$. In addition, $A_{1\mu}^a$ and $A_{2\mu}^a$ $(a=1,2,3)$ are the fields from $SU(2)_{LA,LB}$. Finally, $g'$ and $B_{\mu}^3$ are the hypercharge coupling and field, respectively.

Derived from the above, as it is usually carried out, it can be obtained the physical gauge bosons, including those of the SM~\cite{Moats:2012laz}. In this sense, their  masses until the order of $\mathcal{O}(v^2/(f^2+F^2))$ are
\begin{eqnarray}
m_{\gamma}^2&=&0,\\
m_{Z}^2&=&\frac{1}{4}(g^2+g'\,^2)v^2-(g^2+g'\,^2)
\left(2+\frac{3f^2}{f^2+F^2}(s_g^2-c_g^2)\right)\frac{v^4}{48f^2},\\
m^2_W&=&\frac{1}{4}g^2v^2-g^2\left(2+\frac{3f^2}{f^2+F^2}(s_g^2-c_g^2)\right)\frac{v^4}{48f^2},\\
m_{Z'}^2&=&\frac{1}{4}(g_A^2+g_B^2)(f^2+F^2)-\frac{1}{4}g^2v^2
-\left(2g^2+\frac{3f^2}{f^2+F^2}(g^2+g'\,^2)(s_g^2-c_g^2)\right)\frac{v^4}{48f^2},\\
m_{W'}^2&=&\frac{1}{4}(g_A^2+g_B^2)(f^2+F^2)-m_W^2.\label{wpmass}
\end{eqnarray}

\subsection{Fermion sector}
In order to develop the Yukawa interactions, it should be considered that the fermions transform under $SO(6)_A$ or $SO(6)_B$ symmetries~\cite{Schmaltz:2010ac,Moats:2012laz}. Since the fundamental representation of the $SO(6)$ group comprises two $SU(2)_L$ doublets and two singlets, the fermions are arranged as follows
\begin{equation}
 Q^T=\left(-\frac{1}{\sqrt{2}}(Q_{a1}+Q_{b2}),\frac{i}{\sqrt{2}}(Q_{a1}-Q_{b2}),\frac{1}{\sqrt{2}}(Q_{a2}-Q_{b1}),\frac{i}{\sqrt{2}}(Q_{a2}+Q_{b1}),Q_5,Q_6\right),
\end{equation}
where $Q_a=(Q_{a1},Q_{a2})$ and $Q_b=(Q_{b1},Q_{b2})$ are doublets of $SU(2)_L$ with hypercharge $-1/2$ and $1/2$, respectively. In contrast, $Q_5$ and $Q_6$ are singlets under $SU(2)_L\times SU(2)_R\equiv SO(4)$. To identify $Q_a$ as the SM quark doublet with hypercharge $1/6$, the hypercharge operator must be parametrized as~\cite{Schmaltz:2010ac,Moats:2012laz}
\begin{equation}
Y=T^3_R+T_X=Q_{EM}+T^3_L.
\end{equation}
On the other hand, the fermions that transform under the $SO(6)_B$ fundamental representation are~\cite{Schmaltz:2010ac,Moats:2012laz}
\begin{equation}
 (U^c)^T=\left(-\frac{1}{\sqrt{2}}(U^c_{b1}+U^c_{a2}),
 \frac{i}{\sqrt{2}}(U^c_{b1}-U^c_{a2}),\frac{1}{\sqrt{2}}(U^c_{b2}-U^c_{a1}),
 \frac{i}{\sqrt{2}}(U^c_{b2}+U^c_{a1}),U^c_5,U^c_6\right).
\end{equation}
In this case, the SM up-type singlet will be identified as the fifth component of $U^c$.

To minimize the top quark partner masses, and also the radiative corrections to the Higgs mass, it is proposed the ``bestest" structure for the top quark Yukawa coupling that minimizes the mass of the top partners~\cite{Schmaltz:2010ac,Moats:2012laz}, which can be explicitly casted in the following collective Yukawa coupling
\begin{equation}\label{lag-top}
 \mathcal{L}_t=y_1fQ^TS\Sigma SU^c+y_2fQ'^{T}_a\Sigma U^c+y_3fQ^T\Sigma U'^c_5+y_bfq_3^T(-2iT^2_R\Sigma)U_b^c+\textrm{H.\thinspace c.},
\end{equation}
where
\begin{eqnarray}
Q'_a\,^T&=&\frac{1}{\sqrt{2}}(-Q'_{a1},iQ'_{a1},Q'_{a2},iQ'_{a2},0,0),\\
U'_{5}\,^{cT}&=&(0,0,0,0,U'_5\,^c,0).
\end{eqnarray}
Here, $S$ is the $SO(6)$ matrix $S=diag(1,1,1,1,-1,-1)$, whose inclusion does not break any of the gauge symmetries.

By extracting the physical fields, as it is usually performed, the corresponding masses up to the order of $\mathcal{O}(v^2/f^2)$ are~\footnote{For a detailed presentation of the BLH Yukawa sector see Ref.~\cite{Moats:2012laz}.}
\begin{eqnarray}
m_t^2&=&y_t^2v_1^2,\label{mass-f1}\\
m_b^2&=&y_b^2v_1^2-\frac{2y_b^2}{3\sin^2\beta}\frac{v_1^4}{f^2},\label{mass-f2}\\
m_{T}^2&=&(y_1^2+y_2^2)f^2+\frac{9v_1^2y_1^2y_2^2y_3^2}{(y_1^2+y_2^2)(y_2^2-y_3^2)},\label{mass-f3}\\
m_{T^{5}}^2&=&(y_1^2+y_3^2)f^2-\frac{9v_1^2y_1^2y_2^2y_3^2}{(y_1^2+y_3^2)(y_2^2-y_3^2)},\label{mass-f4}\\
m_{B}^2&=&(y_1^2+y_2^2)f^2,\label{mass-f5}\\
m_{T^{6}}^2&=&m_{T^{2/3}}^2=m_{T^{5/3}}^2=y_1^2f^2\label{mass-f6}.
\end{eqnarray}

\subsubsection{Currents sector}
The Lagrangian that comprises the fermion kinetic terms along with their interactions with gauge bosons is written as follows
\begin{equation}\label{ffv-lag}
\mathcal{L}_V=iQ^{\dagger}\bar{\tau}^{\mu}D_{\mu}Q+iQ'^{\dagger}_a\bar{\tau}^{\mu}D_{\mu}Q'_a
-iU^{c\dagger}\tau^{\mu}D_{\mu}U^c-iU'^{c\dagger}_5\tau^{\mu}D_{\mu}U'^c_5
-iU^{c\dagger}_b\tau^{\mu}D_{\mu}U^c_b,
\end{equation}
where $\tau^\mu$ are the Pauli matrices, $Q$'s and $U$'s are defined as before. The covariant derivatives of the fermion fields, $D_{\mu}Q$'s and $D_{\mu}U$'s, contain the gauge fields of the BLH model, and are defined as~\cite{Martin:2012kqb,Moats:2012laz}.
\begin{eqnarray}
 D_{\mu}Q&=&\partial_{\mu} Q+ig_AA_{1\mu}^aT^aQ+ig'B_{\mu}^3\left({T'}^3+\frac{2}{3}\mathds{1}_{6\times6}\right)Q,\\
 D_{\mu}{Q'}_a&=&\partial_{\mu} Q'_a+ig_AA_{1\mu}^aT^aQ'_a+\frac{ig'}{6}B_{\mu}^3Q'_a,\\
 D_{\mu}U^c&=&\partial_{\mu}U^c+ig_AA_{2\mu}^aT^aU^c+ig'B_{\mu}^3
 \left({T'}^3-\frac{2}{3}\mathds{1}_{6\times6}\right)U^c,\\
 D_{\mu}{U'}^c_5&=&\partial_{\mu}{U'}^c_5-\frac{2ig'}{3}B_{\mu}^3{U'}^c_5,\\
 D_{\mu}U^c_b&=&\partial_{\mu}U^c_b+\frac{ig'}{3}B_{\mu}^3U^c_b.
 \end{eqnarray}

%%%%%%%%%%%%%%%%%%%%%%%%%%%%%%%%%%%%%%%%%%%%%%%%%%%%%%%%%%%%%%%%%%%%%%%%%%%%%%%%%%%%%%%%%%%%%%%%%%%%%%%%%%%%%
%%%%%%%%%%%%%%%%%%%%%%%%%%%%%%%%%%%%%%%%%%%%%%%%%%%%%%%%%%%%%%%%%%%%%%%%%%%%%%%%%%%%%%%%%%%%%%%%%%%%%%%%%%%%%
\section{One-loop amplitudes}\label{OLA}
\begin{center}
\begin{figure}[t!]
\subfloat[]{\includegraphics[width=4.7cm]{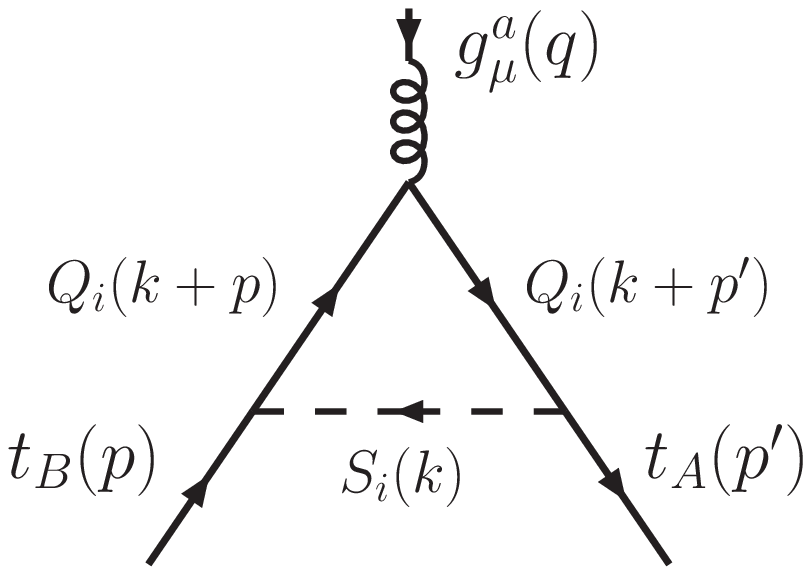}} \qquad
\subfloat[]{\includegraphics[width=4.7cm]{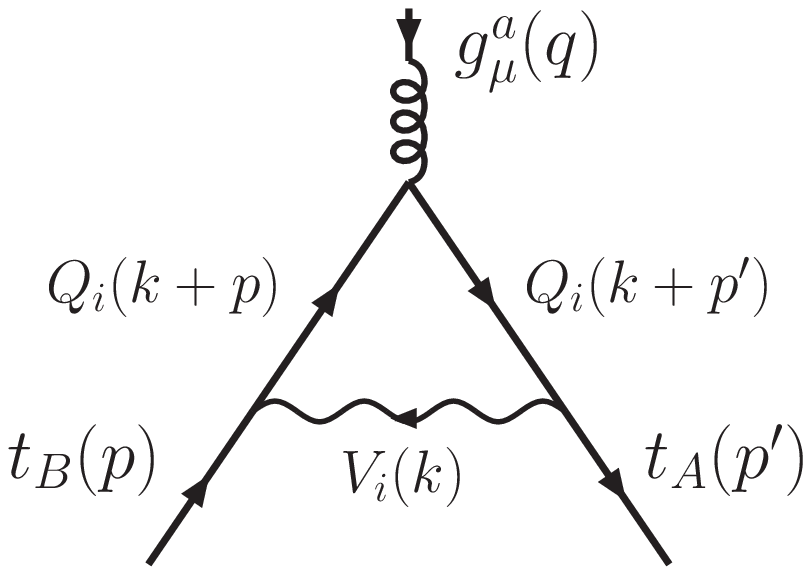}}
\caption{(a) Scalar contributions, $S_i$ $\equiv$ $H^0$, $A^0$, $H'$, $H^{\pm}$, $\phi^0$, $\phi^{\pm}$, $\eta^{0}$, and $\sigma$. (b) Vector contributions, $V_i$ $\equiv$ $Z$, $\gamma$, $W'$, and $Z'$.}
\label{diagramas}
\end{figure}
\end{center}

The generic one-loop diagrams depicted in Fig.~\Ref{diagramas} allow us to construct the amplitudes that involve the BLH contributions to the top quark CMDM and CEDM, $\hat{\mu}_t$ and $\hat{d}_t$, respectively. Regarding to the new physics contributions, the following particles are implicated: $Q_i$ represents the quarks $t$ (SM top quark) $T$, $T^5$, $T^6$, $T^{2/3}$, $T^{5/3}$ and $B$; $S_i$ stands for the scalars $H^0$ (SM Higgs), $A^0$, $H'$, $H^{\pm}$, $\phi^0$, $\phi^{\pm}$, $\eta^{0}$, and $\sigma$; $V_i$ stands for the gauge bosons $Z^0$, $\gamma$, $W'$, and $Z'$. The vertices $\bar{q}q'S_i$ and $\bar{q}q'V_i$ contain the contributions beyond the SM for the $\bar{t}tg$ vertex, which appear in the form factors of the scalar, pseudoscalar, vector and axial structures. In order to construct the corresponding one-loop amplitudes, it is useful to organize the different new contributions to the top quark chromoelectromagnetic dipole moments (CEMDM) as follows
\begin{equation}\label{suma}
\hat{\mu}_t=\sum_i\hat{\mu}_t(S_i)+\sum_i\hat{\mu}_t(V_i).
\end{equation}
These contributions can be obtained from the diagrams depicted in Fig.~\Ref{diagramas}. It should be recalled that we are computing forty four Feynman diagram, in particular, it has thirty one with a virtual scalar $(S_i)$ and twelve with a virtual gauge boson $(V_i)$. The generic amplitude of the virtual scalar boson contributions (see Fig.~\Ref{diagramas}a) has the form
\begin{eqnarray}\label{amplitudesc}
\mathcal{M}_t^{\mu}(S_i) &=& \sum_j\mu^{2\epsilon}\int\frac{d^Dk}{(2\pi)^D}\bar{u}(p')\left(f^\ast_{S_i}+f^\ast_{P_i}\gamma^5\right)
\delta_{A\alpha_1}
\left[i\frac{\slashed{k}+\slashed{p}\thinspace '+m_{Q_j}}{(k+p')^2-m^2_{Q_j}}\delta_{\alpha_1\alpha_3}\right]
\left(-ig_s\gamma^\mu T^a_{\alpha_2\alpha_3}\right)
\nonumber\\
&&\times\left[i\frac{\slashed{k}+\slashed{p}+m_{Q_j}}{(k+p)^2-m^2_{Q_j}}\delta_{\alpha_3\alpha_4}\right]
\left(f_{S_i}+f_{P_i}\gamma^5\right)\delta_{\alpha_4B}u(p)\left(\frac{i}{k^2-m_{S_{i}}^2}\right).
\end{eqnarray}
The $T^a_{\alpha_{k} \alpha_{m}}$ are the $SU(3)_C$ group generators. In contrast, the generic amplitude of the virtual vector boson contributions (see Fig.~\Ref{diagramas}b) can be expressed as
\begin{eqnarray}\label{amplitudvec}
 \mathcal{M}_t^ {\mu}(V_i)&=&\sum_j\mu^{2\epsilon}\int \frac{d^Dk}{(2\pi)^D}\bar{u}(p')\gamma^{a_1}\left(f^\ast_{V_i}+f^\ast_{A_i}\gamma^5\right)\delta_{A\alpha_1}
 \left[i\frac{\slashed{k}+\slashed{p'}+m_{Q_j}}{(k+p')^2-m^2_{Q_j}}\delta_{\alpha_1\alpha_3}\right]
 \left(-ig_s\gamma^{\mu}T^a_{\alpha_2\alpha_3}\right)
\nonumber\\
&&\times\left[i\frac{\slashed{k}+\slashed{p}+m_{Q_j}}{(k+p)^2-m^2_{Q_j}}\delta_{\alpha_3\alpha_4}\right]
\gamma^{a_2}\left(f_{V_i}+f_{A_i}\gamma^5\right)\delta_{\alpha_4B}u(p)
\left[\frac{i}{k^2-m^2_{V_i}}\left(-g_{\alpha_1\alpha_2}+\frac{k_{\alpha_1}k_{\alpha_2}}{m^2_{V_i}}\right)\right].
\end{eqnarray}
For the virtual photon case, the longitudinal term of the propagator is absent. In summary, there are thirty one scalar Feynman diagramas (see Fig.~\ref{diagramas}(a)) and twelve vector Feynman diagrams (see Fig.~\ref{diagramas}(b)), which together give rise to the CEMDM of the top quark. The Feynman rules employed in our calculation are listen in Appendix~\ref{Appendix}.

The CEMDM Lagrangian is defined as \cite{Haberl:1995ek,Khachatryan:2016xws,Bernreuther:2013aga}
\begin{eqnarray}\label{lagrangian}
\mathcal{L}\mathrm{eff} &=&-\frac{1}{2}\bar{q}_A\sigma^{\mu\nu}\left(\mu_q+id_q\gamma^5\right)q_{B}G_{\mu\nu}^aT_{AB}^a,
\end{eqnarray}
where $T_{AB}^a$ represents the color generator of $SU(3)_C$, $A$ and $B$ symbolize color indices,
$\sigma^{\mu\nu}\equiv \frac{i}{2}[\gamma^\mu,\gamma^\nu]$,
$\mu_q$ is the CP-conserving chromomagnetic form factor, $d_q$ is the CP-violating chromoelectric form factor, and $G_{\mu\nu}^a=\partial_\mu g_\nu^a-\partial_\nu g_\mu^a-g_sf_{abc}g_\mu^bg_\nu^c$ is the gluon strength field. In the context of the SM, the CMDM is induced at the one-loop level \cite{Martinez:2007qf,Choudhury:2014lna}, while the CEDM arises at the three-loop level \cite{Czarnecki:1997bu}. In addition, it is usual in the literature to define dimensionless dipoles for quarks \cite{Aranda:2020tox,Haberl:1995ek,PDG2020,Khachatryan:2016xws,Bernreuther:2013aga}, just as
\begin{equation}\label{}
\hat{\mu}_{q}\equiv \frac{m_q}{g_s}\mu_q \quad , \quad \hat{d}_q\equiv \frac{m_q}{g_s}d_q ~,
\end{equation}
where $m_q$ is the quark mass, $g_s=\sqrt{4\pi\alpha_s}$ is the QCD group coupling constant, where $\alpha_s$ represents the strong coupling constant characterized at the mass of the $Z$ gauge boson, being $\alpha_s(m_Z^2)=0.1179$ \cite{PDG2020}.
%%%%%%%%%%%%%%%%%%%%%%%%%%%%%%%%%%%%%%%%%%%%%%%%%%%%%%%%%%%%%%%%%%%%%%%%%%%%%%%%%%%%%%%%%%%%%%%%%%%%%%%%%%%%%%
%%%%%%%%%%%%%%%%%%%%%%%%%%%%%%%%%%%%%%%%%%%%%%%%%%%%%%%%%%%%%%%%%%%%%%%%%%%%%%%%%%%%%%%%%%%%%%%%%%%%%%%%%%%%%%
\section{Phenomenological discussion}\label{Phenomenology}
\subsection{Feynman rules}
As mentioned above, there are several tens of interactions coming from the BLH model that are involved in the calculation of the CMDM (for more details see the Appendix~\ref{Appendix}). Accordingly, in our predictions, we will use the $W$ boson mass and the coupling constants $g'$ and $g$, related to the $U(1)$ and $SU(2)$ gauge groups, as input values. In contrast, the gauge couplings $g_A$ and $g_B$, associated with the $SU(2)_{LA}$ and $SU(2)_{LB}$ gauge bosons, must be determined.

In the Refs.~\cite{Kalyniak:2013eva,Martin:2012kqb,Moats:2012laz} the parameter space of the BLH model has been constrained from ATLAS and CMS Higgs measurements. Among them we can find that $y_t$ is the Yukawa coupling of the top quark, the $B_{\mu}$ parameter that is generated in the collective symmetry breaking, $\lambda_0$ is the Higgs quartic coupling, $\alpha$ is the mixing angle of the mass eigenstates \cite{Moats:2012laz}, and $\tan\beta=v_1/v_2$. The coupling parameters of the BLH model are achieved by the following relations~\cite{Schmaltz:2010ac}
\begin{equation}
g=\frac{g_Ag_B}{\sqrt{g_A^2+g_B^2}},\label{couplingg}
\end{equation}
\begin{equation}
 s_g\equiv\sin\theta_g=\frac{g_A}{\sqrt{g_A^2+g_B^2}},\hspace{0.5cm}c_g\equiv\cos\theta_g=\frac{g_B}{\sqrt{g_A^2+g_B^2}}.
\end{equation}
where $s_g$ and $c_g$ can be computed under the restriction $g=0.6525$ and assuming different values for $g_A$ and $g_B$ in a perturbative approach~\cite{Moats:2012laz}.

\subsection{Parameter space}\label{parametersp}
In order to evaluate the CEMDM of the top quark, we present a summary of the involved parameter space of the BLH model. The parameter space that we will consider in this analysis is generated assuming as an input value the mass of the pseudoscalar $A^0$, which is fixed around $1$ TeV, in strict agreement with the most recent experimental data on searches for new scalar particles~\cite{SNSB}. Additionally, due to perturbative requirements, theoretical constraints on the BLH parameters are imposed~\cite{Schmaltz:2010ac,Kalyniak:2013eva}, such as the mixing angle $\beta$, which is restricted to be
\begin{align}\label{tanb}
1\lesssim \text{tan}\ \beta < \sqrt{ \frac{2+2 \sqrt{\big(1-\frac{m^{2}_{H_0} }{m^{2}_{A_0}} \big) \big(1-\frac{m^{2}_{H_0} }{4 \pi v^{2}}\big) } }{ \frac{m^{2}_{H_0}}{m^{2}_{A_0}} \big(1+ \frac{m^{2}_{A_0}- m^{2}_{H_0}}{4 \pi v^{2}}  \big) } -1 }.
\end{align}
From this inequality we can extract values for the $\beta$ parameter. In particular, by considering that $m_{A^0}=1$ TeV, it can be obtained that $\beta=1.24$ is in good agreement with Eq.~(\ref{tanb}).

By using that $m_{H^0}=125.1$ GeV and $v=246$ GeV~\cite{PDG2020}, the rest of the BLH parameters can be computed~\cite{Schmaltz:2010ac,Kalyniak:2013eva}:
\begin{align}\label{parametros}
B_\mu &=\frac{1}{2}(\lambda_0  v^{2} + m^{2}_{A_{0}}  )\, \text{sin}\, 2\beta,\\
  \lambda_0 &= \frac{m^{2}_{H_{0}}}{v^{2}}\Big(\frac{  m^{2}_{H_{0}}- m^{2}_{A_{0}} }{m^{2}_{H_{0}}  -m^{2}_{A_{0}} \text{sin}^{2}\, 2\beta  }\Big),\\
  \text{tan}\, \alpha &= \frac{ B_\mu \text{cot}\, 2\beta+ \sqrt{(B^{2}_\mu/\text{sin}^{2}\, 2\beta)-2\lambda_0 B_\mu v^{2} \text{sin}\, 2\beta+ \lambda^{2}_{0} v^{4}\text{sin}^{2}\, 2\beta  }  }{B_\mu -\lambda_0 v^{2} \text{sin}\, 2\beta}.\label{alpha}
\end{align}
It should be noted that our input values satisfy the restriction $\lambda_0<4\pi$~\cite{Schmaltz:2010ac}. Tentatively, several other options exist for this parameter space since $m_{A^0}$ would be greater than $800$ GeV, however, our choice is consistent with the current search results for new scalar bosons~\cite{SNSB}. In this way, we find that $m_{H^{\pm}}\sim 1$ TeV and $m_{H^{\prime}}\sim 1.2$ TeV. Besides, by using the phenomenological approach from Ref.~\cite{Kalyniak:2013eva} for $m_\sigma$ (see Eq.~(\ref{msigma})), where
\begin{align}\label{cotaK}
1 < K_\sigma <\frac{16 \pi^{2}}{\lambda_0 (8\pi -\lambda_0)},
\end{align}
it can be derived that $m_\sigma\sim 4$ TeV.

Finally, as it was established in Ref.~\cite{Kalyniak:2013eva}, we fixed the mass of the rest of scalar particles as $m_4=m_5=m_6=30$ GeV. Notice that even though $m_{\eta^0}=m_4$, according to the BLH model, the restriction $m_4\gtrsim 10$ GeV must be considered. On the other hand, $m_4$ is also a free parameter, so the larger its mass is, the more its decoupling effect. In addition, its coupling constant with top quark pairs is very small, being its phenomenological effects negligible.

\subsubsection{Fine-tuning}\label{finet}
Typically, the measure of fine tuning is strongly correlated with the stability of the electroweak Higgs VEV under radiative corrections. In the BLH model, the fine tuning can be computed in the following way~\cite{Schmaltz:2010ac,Moats:2012laz}
\begin{equation}
\Psi=\left|\frac{\delta v^2_{EW}}{v^2_{EW}}\right|=\dfrac{|\delta m^{2}_{1}|}{\lambda_{0} v^{2} \cos^{2} \beta  },
\end{equation}
where $\delta v^2_{EW}$ represents the deviations from $v^2_{EW}$ coming from radiative corrections. Moreover, in the BLH model $\delta m^{2}_{1}$ can be written as
\begin{equation}
\delta m^{2}_{1}=-\frac{27 f^{2}}{8 \pi^{2}} \frac{ y^{2}_{1}y^{2}_{2}y^{2}_{3} }{y^{2}_{2} - y^{2}_{3}} \log\left(\frac{y^{2}_{1} + y^{2}_{2} }{ y^{2}_{1}+ y^{2}_{3}}\right).
\end{equation}
Particularly, when $\Psi\sim 1$ the BLH model does not demand fine tuning. In this sense, a fit on the Yukawa coupling parameters $y_1, y_2, y_3$ is required. In accordance with a perturbative scenario, each one of these $y_i$ parameters should be less than the unity. On the other hand, in the context of the BLH model, the Yukawa coupling of top quark is determined by
\begin{equation}
y_t=\frac{m_t}{v \sin\beta}=3\frac{y_1 y_2 y_3}{\sqrt{(y^2_1+y^2_2)(y^2_1+y^2_3)}},
\end{equation}
where $m_t=172.76$ GeV is the top quark mass~\cite{PDG2020}. Therefore, by evaluating $y_t$ we can randomize perturbative values for the $y_i$ parameters for $i=1,2,3$ under the restriction that $y_t=0.742$. Thus, in order to estimate the fine-tuning measure, we choose the following values: $y_1=0.35$, $y_2=0.47$, and $y_3=0.65$, being $\Psi\sim 1$ when the new physics scale $f$ lies in the interval $[2,4]$ TeV.

Regarding to the $F$ energy scale, by taking into account that $F>f$~\cite{Schmaltz:2010ac} and resorting to Eqs.~(\ref{wpmass}) and (\ref{couplingg}), if we assume that $F\sim 5$ TeV it implies that $m_{W^\prime,Z^\prime}\sim 4$ TeV, which is in agreement with the most recent searches for charged heavy gauge bosons~\cite{CMS:2018hff,ATLAS:2018ihk,CMS:2018fza,ATLAS:2019lsy}.
%%%%%%%%%%%%%%%%%%%%%%%%%%%%%%%%%%%%%%%%%%%%%%%%%%%%%%%%%%%%%%%%%%%%%%%%%%%%%%%%%%%%%%%%%%%%%%%%%%%%%%
\subsection{Numerical results}
As it occurs in the SM, we found that in the BLH model the one-loop contributions to the SM quarks CEDM are also zero. In particular, by taking into account scalar contributions to $\hat{d}_t$, coming from Eq. (\ref{amplitudesc}), where $\hat{d}_t$ depends on the squared moment of the gluon ($q^2$), we obtain a generic formula for each scalar particle
\begin{eqnarray}\label{cedm-escalar}
\hat{d}_t(S_i) &=& \kappa_i(f^{\prime e}_{iP}f^{e}_{iS}+f^{e}_{iP}f^{\prime e}_{iS}).
\end{eqnarray}
Here, $\kappa_i$ is a complex number, $f^{e,\prime e}_{iP,iP}$ and $f^{e,\prime e}_{iS,iS}$ are form factors that depend on $q^2$, the symmetry breaking scale $f$, and the masses of the particles involved (see Fig.~\ref{diagramas}(a)). After solving the four-dimensional integrals and algebraically manipulating the quantity $f^{\prime e}_{iP}f^{e}_{iS}+f^{e}_{iP}f^{\prime e}_{iS}$, it is found analytically that $f^{\prime e}_{iP}f^{e}_{iS}+f^{e}_{iP}f^{\prime e}_{iS}=0$, therefore, $\hat{d}_t$ is zero. This same pattern occurs when
$q^2=0$, that is to say, $\hat{d}_t=0$. As far as vector contributions are concerned, from Eq.~(\Ref{amplitudvec}) and assuming that $q^2\neq0$, the CEDM of the top quark can be expressed as
\begin{eqnarray}\label{cedm-vectorial}
\hat{d}_t(V_i) &=& \kappa'_i(f^{\prime e}_{iV}f^{e}_{iA}-f^{e}_{iV}f^{\prime e}_{iA}).
\end{eqnarray}
For this case, once again $f^{\prime e}_{iV}f^{e}_{iA}-f^{e}_{iV}f^{\prime e}_{iA}=0$ when both $q^2\neq0$ and $q^2=0$, therefore, its contributions to $\hat{d}_t$ are all zero. Here, $\kappa'_i$ is also a complex number and $f^{e,\prime e}_{iA,iA}$ and $f^{e,\prime e}_{iV,iV}$ are form factors that depend on $q^2$, $f$ and the masses of the particles involved (see Fig.~\ref{diagramas}(b)).

By contrast, for the CMDM case, we have two similar structures, for both $q^2\neq0$ and $q^2=0$. The scalar part can be written as
\begin{eqnarray}\label{cmdm}
\hat{\mu}_t(S_i)=\zeta_i (f^{\prime m}_{iP}f^{m}_{iS}-f^{m}_{iP}f^{\prime m}_{iS}),
\end{eqnarray}
whereas the vector one is
\begin{eqnarray}\label{cmdm}
\hat{\mu}_t(V_i)=\zeta'_i (f^{\prime m}_{iV}f^{m}_{iA}-f^{m}_{iV}f^{\prime m}_{iA}),
\end{eqnarray}
where $\zeta,\zeta'$ represent complex numbers; $f^{m,\prime m}_{iP,iP}$, $f^{m,\prime m}_{iS,iS}$, $f^{m,\prime m}_{iA,iA}$, and $f^{m,\prime m}_{iV,iV}$ are also form factors that depend on $q^2$, $f$ and the masses of the particles involved (see Fig.~\ref{diagramas}). Because more than forty Feynman diagrams are contributing to the CMDM of the top quark in the BLH model only numerical results are presented. The numerical evaluation of the top quark CMDM also requieres to determine the masses of the exotic quarks; this is achieved by assuming that $f$ goes from $2$ up to $4$ TeV and inserting it together $y_i$ (see Section~\ref{finet}) in Eqs.~(\ref{mass-f3}-\ref{mass-f6}).
\begin{center}
\begin{figure}[!ht]
\subfloat[]{\includegraphics[width=9.5cm]{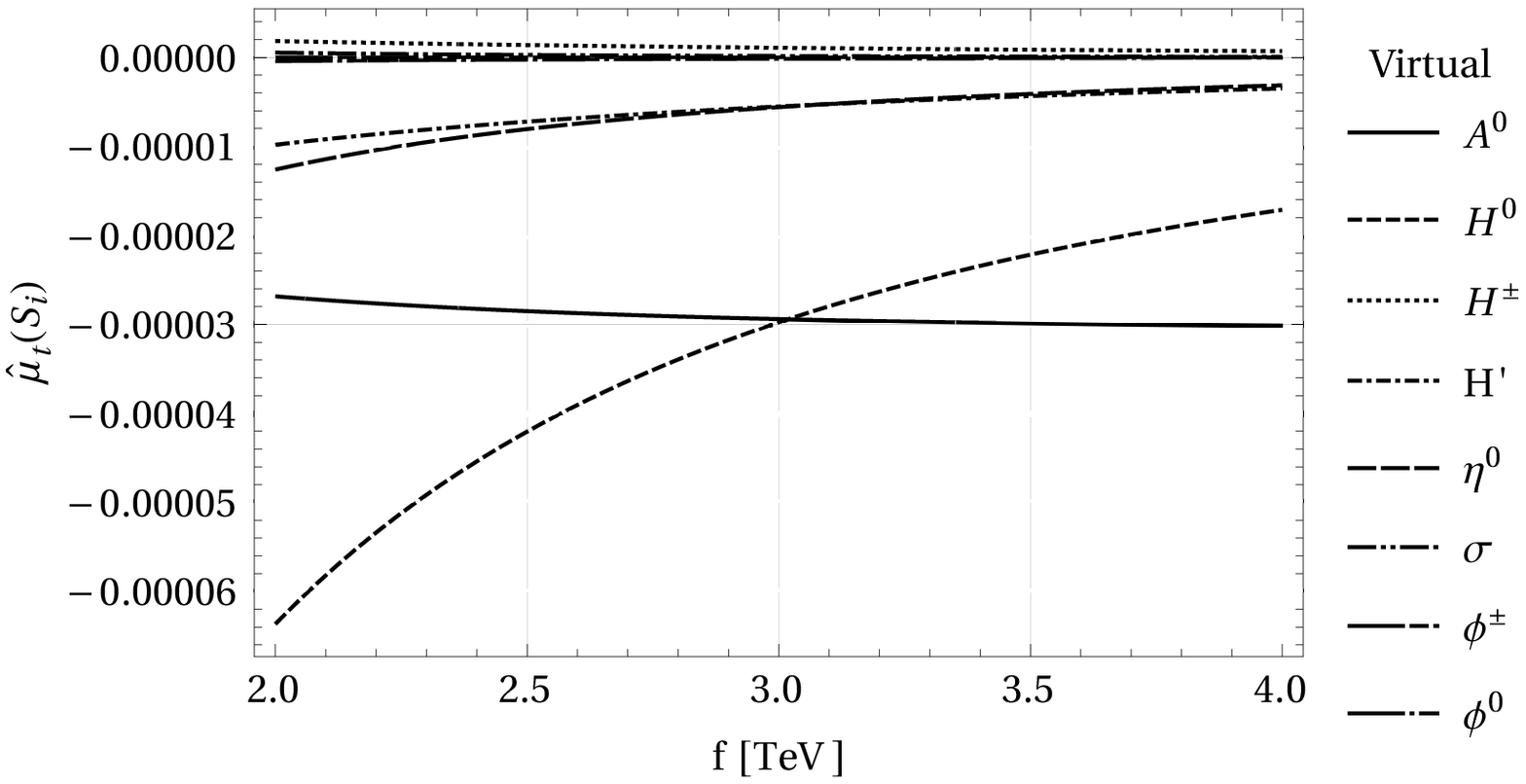}} \quad
\subfloat[]{\includegraphics[width=8.0cm]{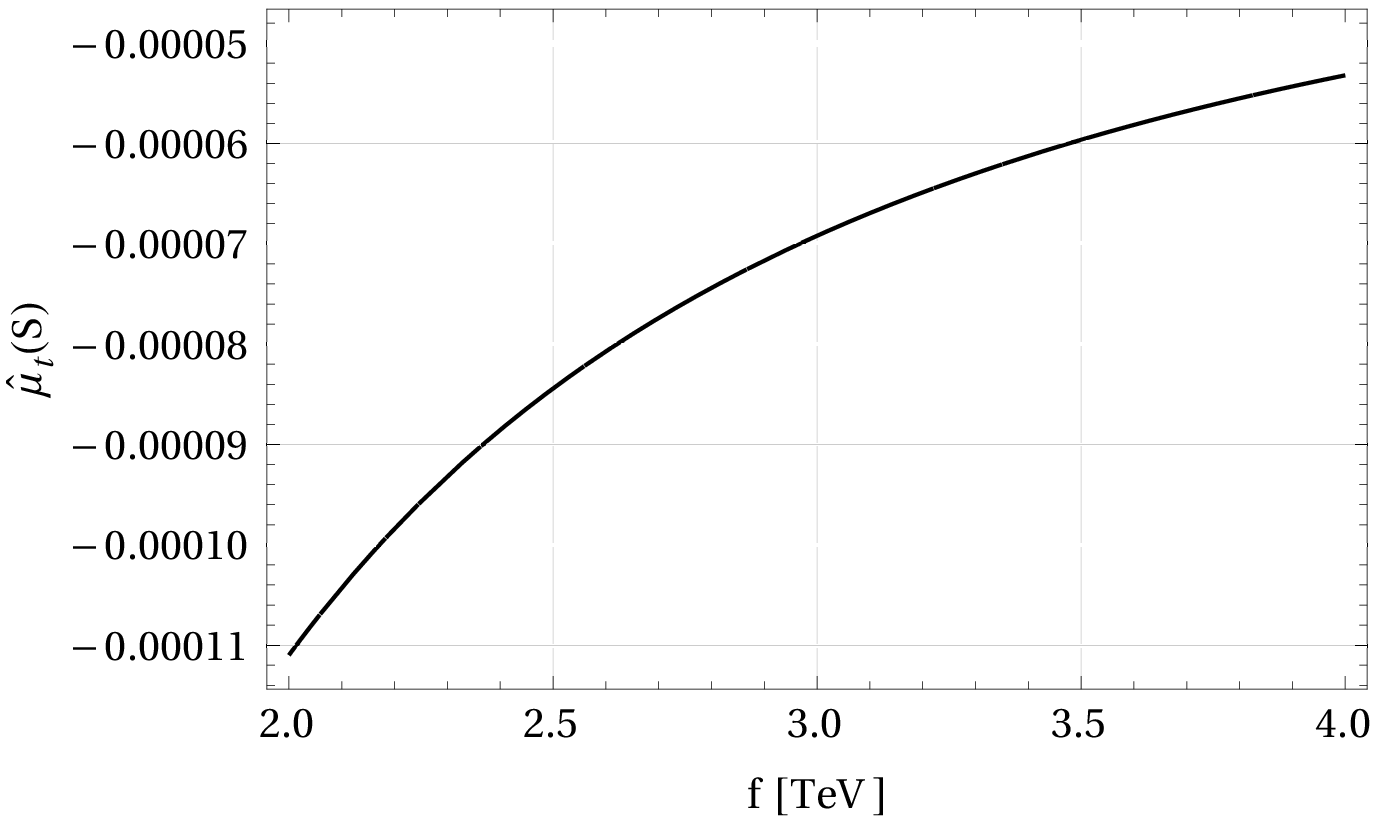}}
\caption{
(a) Individual scalar contributions to $\hat{\mu}_t$ in the BLH model.
(b) Total scalar contribution to $\hat{\mu}_t$ in the BLH model.}
\label{scalars}
\end{figure}
\end{center}
Since the main goal of this work is to study the impact of the exotic particles, predicted in the context of the BLH model, on the top quark CMDM, we will show the behavior of this observable in terms of the new physics scale $f$. In Fig. \ref{scalars} all the contributions to the top quark CMDM in the BLH model, coming from virtual scalar particles, are displayed. In Fig.~\Ref{scalars}(a) it can be appreciated that the most important contributions are due to $H$ and $A^0$. Therefore, as it can be seen in Fig.~\Ref{scalars}(b), the dominant contributions essentially determine the behavior of $\hat{\mu}_t(S)$, where the sum of all scalar contributions are being presented. It should be recalled that the analysis interval for $f$ goes from 2 up to 4 TeV because of the restriction for the fine-tuning parameter (see Section~\ref{finet}).
\begin{center}
\begin{figure}[!ht]
\subfloat[]{\includegraphics[width=9.5cm]{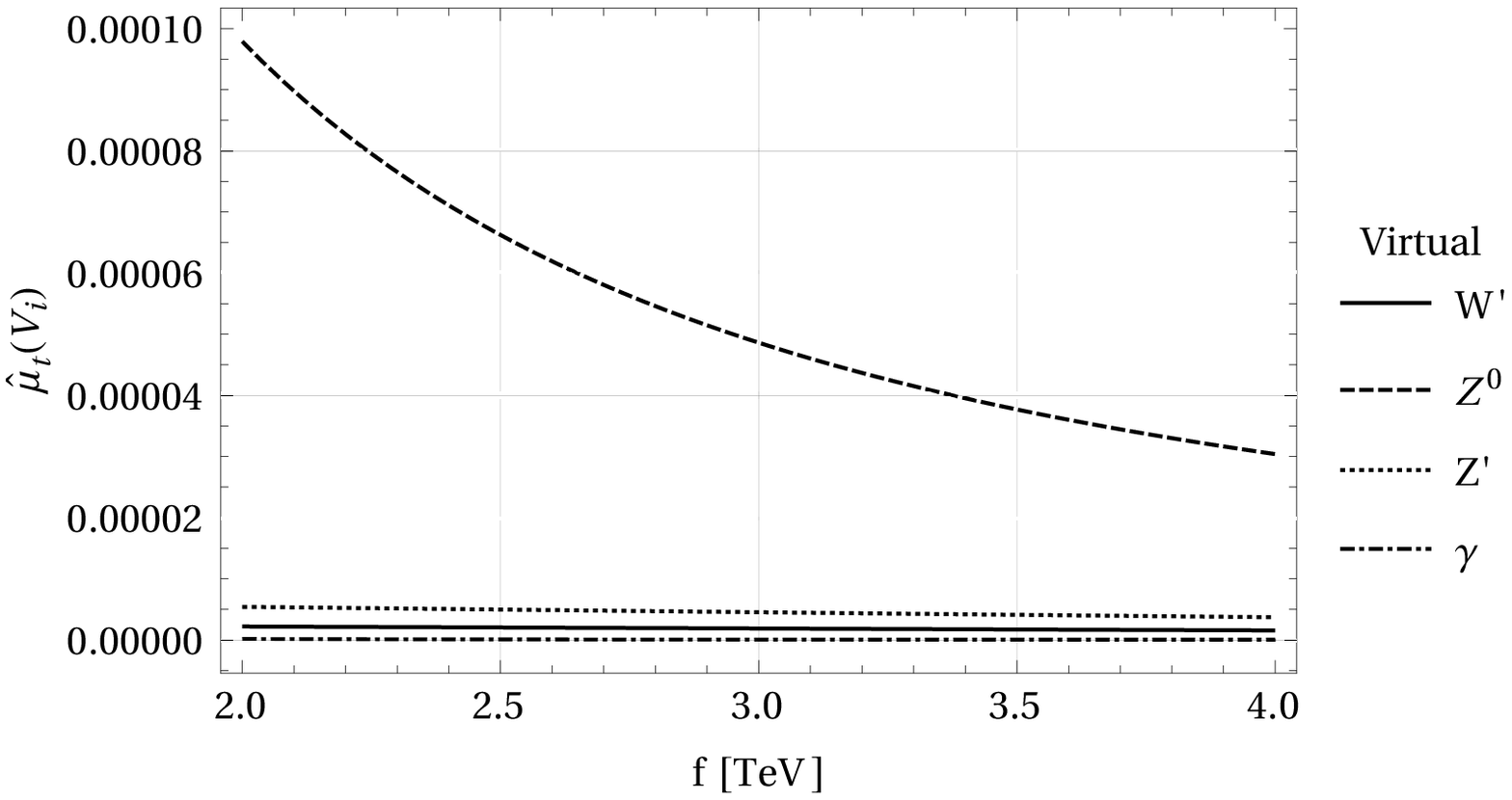}}
\quad
\subfloat[]{\includegraphics[width=8.0cm]{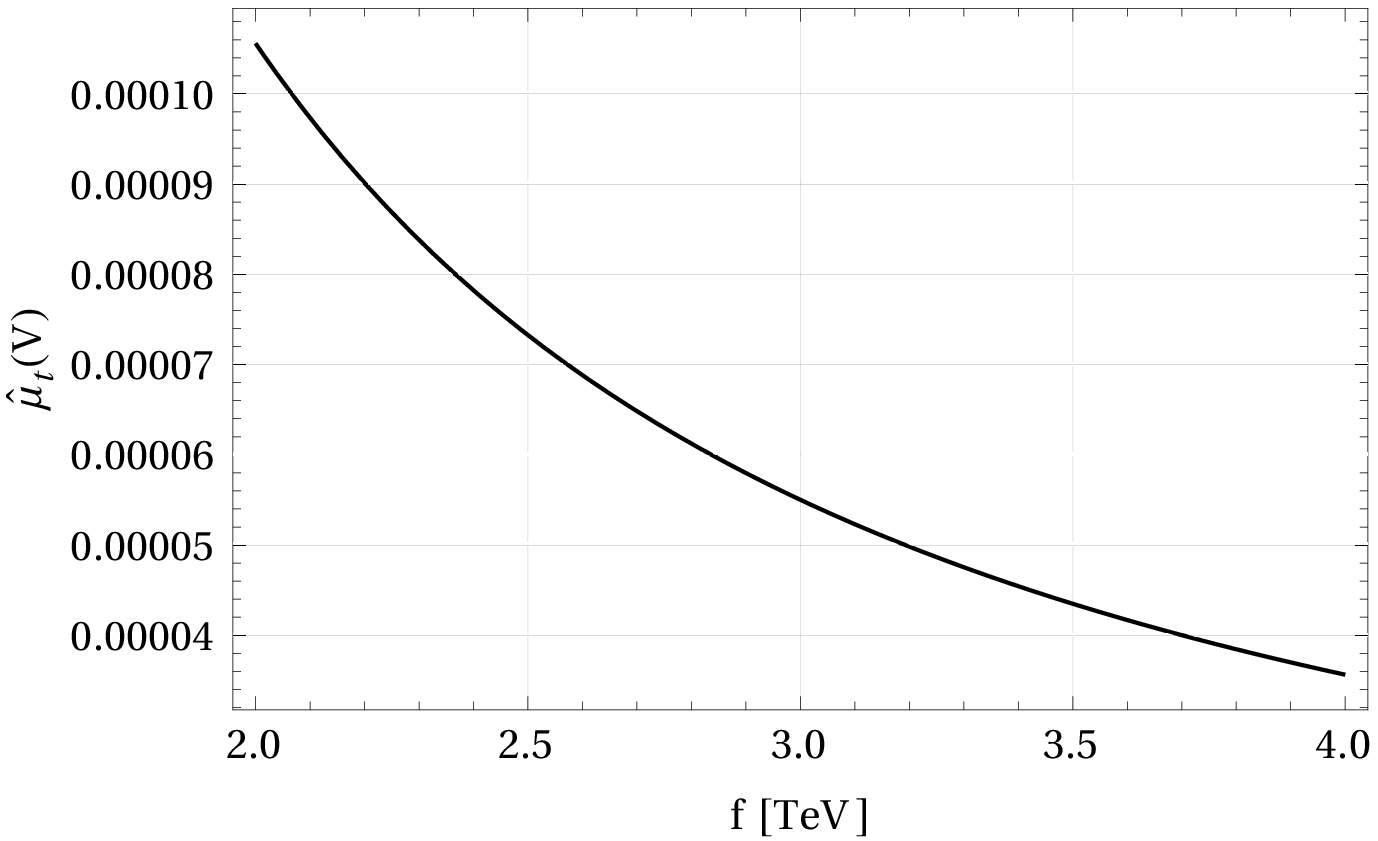}}
\caption{
(a) Individual vector contributions to $\hat{\mu}_t$ in the BLH model.
(b) Total vector contribution to $\hat{\mu}_t$ in the BLH model.}
\label{e-vectorial}
\end{figure}
\end{center}
Now, in Fig.~\ref{e-vectorial} it is shown each contribution of the virtual gauge bosons to the CMDM in the BLH model. Here, the $Z^0$ gauge boson provides the most important signal.
It is remarkable that $|\hat{\mu}_t(S)|$ is slightly more intense than $|\hat{\mu}_t(V)|$, hence the scalar part of $\hat{\mu}_t$ will be the most relevant, rendering it a negative quantity.
\begin{center}
\begin{figure}[!ht]
\includegraphics[width=9cm]{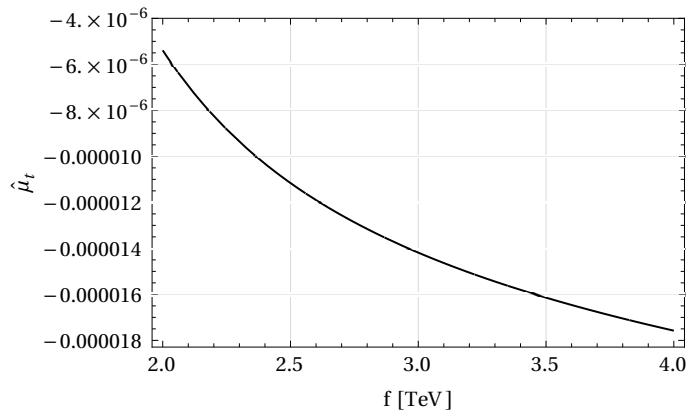}
\caption{Total contribution to $\hat{\mu}_t$ in the BLH model.}
\label{suma-total}
\end{figure}
\end{center}
In Fig.~\Ref{suma-total} the total contribution (all scalar and vector parts) to $\hat{\mu}_t$ in the BLH model is appreciated. Our predictions for $|\hat{\mu}_t|$ ranges between $10^{-6}$ and $10^{-5}$ for $f=2$ TeV and $f=4$ TeV, respectively. Within the Little Higgs models, we can compare our predictions with the results previously obtained in the framework of the Littlest Higgs model with T-parity, where it is established that $|\hat{\mu}_t|$ would be of the order of $10^{-4}$~\cite{Cao:2008qd}, for a new particle mass scale $f=500$ GeV. The authors of Ref.~\cite{Cao:2008qd} does not present results at the TeV scale, however, it can be inferred that its prediction would be much less intense due to the decoupling effects. Furthermore, in order of the top pair production process take place, if we focus on the energy transfer through $q\bar q\to g\to t\bar t$, it is observed from their predictions that the intensity of $|\hat{\mu}_t|$ would decrease by an order of magnitude if the invariant mass rounds units of TeV.

\section{Conclusions}\label{Conclusions}
We have calculated the CMDM of the top quark, induced at the one-loop level, in the context of the BLH model. The BLH model, besides offering a solution to the hierarchy problem of the SM, is also attractive since it preserves the custodial symmetry together with the absence of fine tuning. This extended model, in agreement with recent studies of its parameter space, predicts new degrees of freedom whose masses would be just at a few TeVs, which could be of great interest for both experimental, inside the Large Hadron Collider, and phenomenological testings.

Our BLH parameter space is motivated by considering a pseudoscalar boson mass around 1 TeV (in accordance with the most recent bounds on the heavy scalar masses), the absence of fine tuning, and a perturbative approach for the coupling constants of the BLH model. This implies that the new breaking scale $f$ would be between $2$ TeV and $4$ TeV. Based on this energy scale interval, our predictions for the CMDM of the SM top quark are of the order of $10^{-6}-10^{-5}$, where the dominant contributions correspond to virtual scalar particles that circulate in the loops, being the main part that concerning to the SM Higgs boson. Therefore, the extended Yukawa sector of the BLH model is very important in terms of phenomenology for this observable. It is also important to stress that this work provides a summary of new Feynman rules for the BLH model, which have not been previously reported in the literature.

\section*{Acknowledgments}
This work has been partially supported by CONACYT, SNI-CONACYT,
 and CIC-UMSNH. J. M. D. thanks C\'atedras CONACYT project 1753.

\appendix
\section{Feynman rules in the BLH model}\label{Appendix}
In the following, we present the used Feynman rules that induce the top quark CMDM in the BLH framework. In order to reduce the writing of Feynman rules, it is introduced some useful definitions:
\begin{align}
 Y_1 &=\sqrt{y_1^2+y_2^2}, & Y_2 &=\sqrt{y_1^2+y_3^2}, \nonumber &\\
 Y_3 &=2y_1^2-y_3^2, & Y_4 & =y_1^2+y_3^2,\nonumber &\\
Y_5 &=y_2^2+3y_3^2, & Y_6 &=\sqrt{f^2+F^2}. \nonumber
\end{align}

\begin{table}[H]
\centering
\renewcommand{\arraystretch}{2.2}
\resizebox{16cm}{!} {
\begin{tabular}{|c|p{15cm}|}
\hline
\textbf{Vertex} & \textbf{Feynman rules} \\
\hline
$A^0t\bar{t}$  & $  \dfrac{-ic_{\beta}(P_L + P_R)y_2y_3(4m_Ws_{\beta}Y_3Y_2 +
    3fgy_1Y_4)}{2fgY_1Y_4^{3/2}} $ \\
\hline
$A^0t\bar{T}$  & $  \dfrac{-ic_{\beta}}{fg(y_1^2 + y_2^2)^{3/2}Y_4^{3/2}}\bigg(2fgP_R(2y_1^4 + y_1^2y_2^2 - y_2^4)y_3Y_4 +
    2m_Ws_{\beta}y_1Y_2 \>
    \Big(-2P_Ry_3(-y_1^4 + y_2^4 + y_1^2y_3^2 + y_2^2y_3^2)- P_Ly_2\big(2y_1^4 + 2y_2^2y_3^2 + y_1^2(2y_2^2 + 11y_3^2)\big)\Big)\bigg) $ \\
\hline
$A^0t\bar{T}^5$  & $ \dfrac{ic_{\beta}}{2fgY_1Y_4^{3/2}(-y_2^2 + y_3^2)}\bigg(fgP_Ly_2(y_2^2 - y_3^2)(2y_1^4 + y_1^2y_3^2 - y_3^4) + m_WP_Rs_{\beta}y_1Y_1y_3\Big(2y_2^2y_3^2
+ 4y_3^4 -
      y_2^8y_3^4Y_2 + y_2^6y_3^6Y_2 +
      2y_1^{10}(y_2^2 - y_3^2)Y_2 + y_1^8Y_2
       (3y_2^4 + y_2^2y_3^2 - 4y_3^4)
    - 2y_1^6y_3^2Y_2
       (-3y_2^4 + 2y_2^2y_3^2 + y_3^4) + y_1^4y_2^2Y_2
       (-y_2^6 + y_2^4y_3^2 + 3y_2^2y_3^4 - 3y_3^6)
     -2y_1^2(8y_2^2 - 2y_3^2 + y_2^8y_3^2Y_2 -
        y_2^6y_3^4Y_2)\Big)\bigg) $ \\
\hline
$  A^0t\bar{T}^6 $  & $ \dfrac{iy_1}{2fg}\bigg(\frac{2fgP_Ls_{\beta}y_2}{Y_1} +
    m_WP_Ry_3\Big(\dfrac{8c_{\beta}^2}{Y_2} +
      s_{\beta}^2\big(2y_1^8  - y_1^2y_2^6
      + 3y_1^4y_2^2y_3^2
      - y_2^6y_3^2 - \dfrac{4}{Y_2} + y_1^6(3y_2^2 + 2y_3^2)\big)\Big)\bigg) $ \\
\hline
$A^0t\bar{T}^{23} $ &  $ \dfrac{ic_{\beta}}{fgY_1Y_4^{3/2}}\bigg(fgP_Ry_1Y_1y_3Y_4 +
    m_Ws_{\beta}Y_2\Big(2P_RY_1y_3(y_1^2 - y_3^2)
    + P_L(-4y_1^3y_2 + 5y_1y_2y_3^2)\Big)\bigg) $ \\
\hline
$ H't\bar{t} $  & $ \dfrac{(P_L + P_R)s_{\alpha}y_2y_3(4m_Ws_{\beta}Y_3Y_2 + 3fgy_1Y_4)}{2fgY_1Y_4^{3/2}}  $  \\
\hline
$  H't\bar{T} $  & $ \dfrac{s_{\alpha}} {2fg(y_1^2 + y_2^2)^{3/2}Y_4^{3/2}}\bigg(fgP_R(2y_1^4 + y_1^2y_2^2 - y_2^4)y_3Y_4 +
    2m_Ws_{\beta}y_1Y_2
     \Big(-2P_Ry_3(-y_1^4
    + y_2^4 + y_1^2y_3^2 + y_2^2y_3^2)  -
      P_Ly_2\big(2y_1^4 - 4y_2^2y_3^2 + y_1^2(2y_2^2 + 5y_3^2)\big)\Big)\bigg)  $  \\
\hline
$  H't\bar{T}^5 $  & $ \dfrac{s_{\alpha}}{2fgY_1Y_4^{3/2}(-y_2^2 + y_3^2)}\bigg(fgP_Ly_2(y_2^2 - y_3^2)(-2y_1^4 - y_1^2y_3^2 + y_3^4) +
    m_WP_Rs_{\beta}y_1Y_1y_3\Big(-10y_2^2y_3^2
      + 4y_3^4
      -y_2^8y_3^4Y_2 + y_2^6y_3^6Y_2 +
      2y_1^10(y_2^2 - y_3^2)Y_2 + y_1^8Y_2
       (3y_2^4 + y_2^2y_3^2 - 4y_3^4)
- 2y_1^6y_3^2Y_2
       (-3y_2^4 + 2y_2^2y_3^2 + y_3^4) + y_1^4y_2^2Y_2
       (-y_2^6 + y_2^4y_3^2 + 3y_2^2y_3^4 - 3y_3^6) +
      y_1^2(8y_2^2 + 4y_3^2 - 2y_2^8y_3^2Y_2 +
        2y_2^6y_3^4Y_2)\Big)\bigg)  $  \\
\hline
$ H't\bar{T}^6  $  & $ -\dfrac{y_1}{fg}\bigg(\dfrac{8c_{\beta}m_WP_Rs_{\alpha}y_3}{Y_2} +
     c_{\alpha}\Big(\frac{2fgP_Ly_2}{Y_1} + m_WP_Rs_{\beta}y_3
        \big(-2y_1^8
        + y_1^2y_2^6 - 3y_1^4y_2^2y_3^2 + y_2^6y_3^2 -
         \frac{4}{Y_2} - y_1^6(3y_2^2 + 2y_3^2)\big)\Big)\bigg)  $  \\
\hline
$  H't\bar{T}^{23} $  & $ \dfrac{1}{fgY_1Y_4^{3/2}}\bigg(fgP_Rs_{\alpha}y_1Y_1y_3Y_4 +
   m_WY_2\Big(2c_{\alpha}c_{\beta}P_Ly_1y_2Y_4
      +s_{\alpha}s_{\beta}\big(P_Ly_1y_2(2y_1^2 - 7y_3^2)+ 2P_RY_1y_3
        (y_1^2 - y_3^2)\big)\Big)\bigg)  $  \\
\hline
$ H^0t\bar{T} $  & $ \dfrac{c_{\alpha}}{2fgY_1^{3/2}Y_4^{3/2}}\bigg(-\big(fgP_R(2y_1^4 + y_1^2y_2^2 - y_2^4)y_3Y_4\big) +
    2m_Ws_{\beta}y_1Y_2\Big(2P_Ry_3(-y_1^4
      + y_2^4 + y_1^2y_3^2 + y_2^2y_3^2) +
      P_Ly_2\big(2y_1^4 - 4y_2^2y_3^2 + y_1^2(2y_2^2 + 5y_3^2)\big)\Big)\bigg)  $  \\
\hline
$ H^0t\bar{T}^5 $  & $ \dfrac{c_{\alpha}}{2fgY_1Y_4^{3/2}(-y_2^2 + y_3^2)}\bigg(fgP_Ly_2(y_2^2 - y_3^2)(2y_1^4 + y_1^2y_3^2 - y_3^4) -
    m_WP_Rs_{\beta}y_1Y_1y_3\Big(
 -10y_2^2y_3^2 + 4y_3^4-
      y_2^8y_3^4Y_2 + y_2^6y_3^6Y_2 +
      2y_1^10(y_2^2 - y_3^2)Y_2 + y_1^8Y_2
       (3y_2^4 + y_2^2y_3^2 - 4y_3^4)
 - 2y_1^6y_3^2Y_2
       (-3y_2^4 + 2y_2^2y_3^2 + y_3^4) + y_1^4y_2^2Y_2
       (-y_2^6 + y_2^4y_3^2 + 3y_2^2y_3^4 - 3y_3^6)
+
      y_1^2(8y_2^2 + 4y_3^2 - 2y_2^8y_3^2Y_2 +
        2y_2^6y_3^4Y_2)\Big)\bigg)  $  \\
\hline
$ H^0t\bar{T}^6 $  & $  -\dfrac{y_1}{fg}\bigg(\dfrac{2fgP_Ls_{\alpha}y_2}{Y_1} +
     m_WP_Ry_3\Big(\dfrac{-8c_{\alpha}c_{\beta}}{Y_2} +
       s_{\alpha}s_{\beta}\big(-2y_1^8 + y_1^2y_2^6
        - 3y_1^4y_2^2y_3^2 + y_2^6y_3^2 -
         \dfrac{4}{Y_2} - y_1^6(3y_2^2 + 2y_3^2)\big)\Big)\bigg) $  \\
\hline
\end{tabular}
}
\caption{First set of Yukawa couplings in the BLH model.}
\end{table}

%%%%%%%%%%%%%%%%%%%%%%%%%%%%%%%%%%%%%%%%%%%%%%%%%%%%%%%%%%%%%%%%%%%%%%%%%%%%%%%%%%%%%%%%%%%%%%%%%%%%%%%%%%%%%

\begin{table}[H]
\centering
\renewcommand{\arraystretch}{2.2}
\resizebox{16cm}{!} {
\begin{tabular}{|c|p{15cm}|}
\hline
\textbf{Vertex} & \textbf{Feynman rules} \\
\hline
$ H^0t\bar{T}^{23} $  & $ \dfrac{1}{fgY_1Y_4^{3/2}}\bigg(2c_{\beta}m_WP_Ls_{\alpha}y_1y_2Y_4^{3/2} -
   c_{\alpha}\Big(fgP_Ry_1Y_1y_3Y_4
   +
     m_Ws_{\beta}Y_2\big(P_Ly_1y_2(2y_1^2 - 7y_3^2) +
       2P_RY_1y_3(y_1^2 - y_3^2)\big)\Big)\bigg)  $  \\
\hline
$ \sigma t\bar{t} $  & $ -\dfrac{3c_{\beta}m_W(P_L + P_R)y_1y_2y_3}{\sqrt{2}fgY_1
   Y_2}  $  \\
\hline
$  \sigma t\bar{T} $  & $ -\dfrac{c_{\beta}m_WP_R(2y_1^2 - y_2^2)y_3}
   {\sqrt{2}fgY_1Y_2)}  $  \\
\hline
$  \sigma t\bar{T}^5 $  & $ \dfrac{c_{\beta}m_WP_Ly_2(2y_1^2 + 5y_3^2)}{\sqrt{2}fgY_1Y_2}  $  \\
\hline
$\sigma t\bar{T}^6  $  & $ \dfrac{\sqrt{2}m_Ws_{\beta}\Big(-\big(P_Ly_1y_2(y_1^2 - 2y_3^2)\big) +
    2P_RY_1y_3Y_4\Big)}{fgY_1Y_4}  $  \\
\hline
$ \sigma t\bar{T}^{23} $  & $ -\dfrac{\sqrt{2}c_{\beta}m_WP_Ry_1y_3}{fgY_2}  $  \\
\hline
$ H^{+} t\bar{T}^{53} $  & $  \dfrac{3\sqrt{2}c_{\beta}m_WP_Ls_{\beta}y_1y_2y_3^2}
  {Y_1(fgy_1^2 + fgy_3^2)} $  \\
\hline
$ \phi^0 t\bar{t} $  & $ \dfrac{3iFm_W(P_L + P_R)s_{\beta}y_1y_2y_3}{2fY_6g
   Y_1Y_2}  $  \\
\hline
$  \phi^0 t\bar{T} $  & $ \dfrac{-iFm_WP_Rs_{\beta}(2y_1^2 - y_2^2)y_3}{2fY_6gY_1Y_2}\bigg(-1 + y_1^8Y_2 +
    y_2^6y_3^2Y_2 + 3y_1^4y_2^2Y_2
     \big(y_2^2 + y_3^2\big)
     + y_1^6Y_2\big(3y_2^2 + y_3^2\big) +
    y_1^2y_2^4Y_2Y_5\bigg)  $  \\
\hline
$ \phi^0 t\bar{T}^5 $  & $ \dfrac{iFm_WP_Ls_{\beta}y_2Y_3}{2fY_6gY_1Y_2}  $  \\
\hline
$ \phi^0 t\bar{T}^6 $  & $  \dfrac{ic_{\beta}Fm_WP_Ly_1y_2}
  {fY_6gY_1}  $  \\
\hline
$ \phi^0 t\bar{T}^{23} $  & $ \dfrac{iFm_WP_Rs_{\beta}y_1y_3}
  {fY_6gY_2}  $  \\
\hline
$ \phi^{+} t\bar{T}^{53}  $  & $\dfrac{-i\sqrt{2}Fm_WP_Rs_{\beta}y_1y_3}
  {fY_6gY_2}  $  \\
\hline
$ \eta^0 t\bar{t} $  & $ \dfrac{-3im_W(P_L + P_R)s_{\beta}y_1y_2y_3}{2fgY_1Y_2}  $  \\
\hline
$ \eta^0 t\bar{T} $  & $   \dfrac{im_WP_Rs_{\beta}(2y_1^2 - y_2^2)y_3}{2fgY_1Y_2}
   \bigg(-1 + y_1^8Y_2 + y_2^6y_3^2Y_2 +
    3y_1^4y_2^2Y_2\big(y_2^2 + y_3^2\big)
    +
    y_1^6Y_2\big(3y_2^2 + y_3^2\big) + y_1^2y_2^4Y_2
     Y_5\bigg) $  \\
\hline
$ \eta^0 t\bar{T}^5 $  & $  \dfrac{-im_WP_Ls_{\beta}y_2Y_3}{2fgY_1Y_2}  $  \\
\hline
$ \eta^0 t\bar{T}^6 $  & $   \dfrac{-ic_{\beta}m_WP_Ly_1y_2}{fgY_1} $  \\
\hline
$ \eta^0 t\bar{T}^{23} $  & $  \dfrac{-im_WP_Rs_{\beta}y_1y_3}{fgY_2} $  \\
\hline
\end{tabular}
}
\caption{Second set of Yukawa couplings in the BLH model.}
\end{table}

%%%%%%%%%%%%%%%%%%%%%%%%%%%%%%%%%%%%%%%%%%%%%%%%%%%%%%%%%%%%%%%%%%%%%%%%%%%%%%%%%%%%%%%%%%%%%%%%%%%%%%%%%%
%%%%%%%%%%%%%%%%%%%%%%%%%%%%%%%%%%%%%%%%%%%%%%%%%%%%%%%%%%%%%%%%%%%%%%%%%%%%%%%%%%%%%%%%%%%%%%%%%%%%%%%%%%

\begin{table}[H]
\centering
\renewcommand{\arraystretch}{2.2}
\resizebox{16cm}{!} {
\begin{tabular}{|c|p{15cm}|}
\hline
\textbf{Vertex} & \textbf{Feynman rules} \\
\hline
$  Zt\bar{T} $  & $  \dfrac{ic_p(-1 + \gamma^5)\gamma_{\mu}s_{\beta}s_Wv(2y_1^2 - y_2^2)y_3}{24fg'(y_1^2 + y_2^2)Y_2}
    \bigg(-3g^2(y_1^2 + y_2^2)^3Y_4^{3/2} + g'^2\Big(-8 + y_1^8Y_2 + y_2^6y_3^2Y_2
    +
       3y_1^4y_2^2Y_2(y_2^2 + y_3^2) + y_1^6Y_2
        (3y_2^2 + y_3^2) + y_1^2y_2^4Y_2Y_5\Big)\bigg) $  \\
\hline
$ Zt\bar{T}^5 $  & $  \dfrac{ic_p\gamma_{\mu}s_{\beta}s_Wv}{12fg'Y_1Y_4^{3/2}}
 \bigg(-3g^2(1 + \gamma^5)y_2Y_3Y_2 +
     g'^2\Big(-2y_1^2\big(-4(-1 + \gamma^5)Y_1y_3
     + 3(1 + \gamma^5)y_2
          Y_2\big) + y_3^2\big(8(-1 + \gamma^5)Y_1y_3 +
         3(1 + \gamma^5)y_2Y_2\big)\Big)\bigg)  $  \\
\hline
$ Zt\bar{T}^6 $  & $   \dfrac{-ic_{\beta}c_p(1 + \gamma^5)(g^2 + g'^2)\gamma_{\mu}s_Wvy_2}
   {2fg'Y_1} $  \\
\hline
\end{tabular}
}
\caption{First set of current sector couplings in the BLH model.}
\end{table}

\begin{table}[H]
\centering
\renewcommand{\arraystretch}{2.2}
\resizebox{16cm}{!} {
\begin{tabular}{|c|p{15cm}|}
\hline
\textbf{Vertex} & \textbf{Feynman rules} \\
\hline
$Zt\bar{T}^{23}  $  & $   \dfrac{ic_p(-1 + \gamma^5)(g^2 + g'^2)\gamma_{\mu}s_{\beta}s_Wvy_3}
   {2fg'Y_2} $  \\
\hline
$ Z't\bar{t} $  & $ \dfrac{ic_pg(1 + \gamma^5)g_A\gamma_{\mu}}{4g_B}  $  \\
\hline
$ Z't\bar{T} $  & $   \dfrac{ic_pg(-1 + \gamma^5)g_B\gamma_{\mu}s_{\beta}v(2y_1^2 - y_2^2)(y_1^2 + y_2^2)^2y_3Y_4}
   {8fg_A} $  \\
\hline
$ Zt\bar{T}^5 $  & $  \dfrac{-ic_pg(1 + \gamma^5)g_A\gamma_{\mu}s_{\beta}vy_2Y_3}
   {4fg_BY_1Y_4}  $  \\
\hline
$ Zt\bar{T}^6 $  & $ \dfrac{-ic_{\beta}c_pg(1 + \gamma^5)g_A\gamma_{\mu}vy_2}{2fg_BY_1}  $  \\
\hline
$Zt\bar{T}^{23}  $  & $ \dfrac{-ic_pg(-1 + \gamma^5)g_B\gamma_{\mu}s_{\beta}vy_3}{2fg_AY_2}  $  \\
\hline
$ \gamma t\bar{T} $  & $ \dfrac{-ig}{6f(y_1^2 + y_2^2)Y_2}\bigg((-1 + \gamma^5)\gamma_{\mu}s_{\beta}s_Wv(2y_1^2 - y_2^2)y_3
    \Big(-2 + y_1^8Y_2 + y_2^6y_3^2Y_2 +
     3y_1^4y_2^2Y_2(y_2^2 + y_3^2) + y_1^6Y_2
      (3y_2^2 + y_3^2) + y_1^2y_2^4Y_2Y_5\Big)\bigg)  $  \\
\hline
$ \gamma t\bar{T}^5 $  & $  \dfrac{-2ig(-1 + \gamma^5)\gamma_{\mu}s_{\beta}s_Wvy_3}{3fY_2} $  \\
\hline
$ {W'}^{+}t\bar{B} $  & $ \dfrac{ig}{4\sqrt{2}fg_A}(-1 + \gamma^5)g_B\gamma_{\mu}s_{\beta}v(2y_1^2 - y_2^2)(y_1^2 + y_2^2)^2y_3Y_4  $  \\
\hline
\end{tabular}
}
\caption{Second set of current sector couplings in the BLH model.}
\end{table}


\section*{References}

\begin{thebibliography}{99}

%\cite{ArkaniHamed:2001nc}
\bibitem{ArkaniHamed:2001nc}
N.~Arkani-Hamed, A.~G.~Cohen and H.~Georgi,
``Electroweak symmetry breaking from dimensional deconstruction,''
Phys. Lett. B \textbf{513} (2001), 232-240
doi:10.1016/S0370-2693(01)00741-9
[arXiv:hep-ph/0105239 [hep-ph]].
%1378 citations counted in INSPIRE as of 22 Sep 2020
%\cite{ArkaniHamed:2002qy}

\bibitem{ArkaniHamed:2002qy}
N.~Arkani-Hamed, A.~G.~Cohen, E.~Katz and A.~E.~Nelson,
``The Littlest Higgs,''
JHEP \textbf{07} (2002), 034
doi:10.1088/1126-6708/2002/07/034
[arXiv:hep-ph/0206021 [hep-ph]].
%1274 citations counted in INSPIRE as of 22 Sep 2020
%\cite{ArkaniHamed:2002pa}

\bibitem{ArkaniHamed:2002pa}
N.~Arkani-Hamed, A.~G.~Cohen, T.~Gregoire and J.~G.~Wacker,
``Phenomenology of electroweak symmetry breaking from theory space,''
JHEP \textbf{08} (2002), 020
doi:10.1088/1126-6708/2002/08/020
[arXiv:hep-ph/0202089 [hep-ph]].
%324 citations counted in INSPIRE as of 22 Sep 2020
%\cite{ArkaniHamed:2002qx}

\bibitem{ArkaniHamed:2002qx}
N.~Arkani-Hamed, A.~G.~Cohen, E.~Katz, A.~E.~Nelson, T.~Gregoire and J.~G.~Wacker,
%``The Minimal moose for a little Higgs,''
JHEP \textbf{08} (2002), 021
doi:10.1088/1126-6708/2002/08/021
[arXiv:hep-ph/0206020 [hep-ph]].
%698 citations counted in INSPIRE as of 22 Sep 2020
%\cite{Schmaltz:2010ac}

\bibitem{Chang:2003zn}
S.~Chang,
``A 'Littlest Higgs' model with custodial SU(2) symmetry,''
JHEP \textbf{12} (2003), 057
doi:10.1088/1126-6708/2003/12/057
[arXiv:hep-ph/0306034 [hep-ph]].
%185 citations counted in INSPIRE as of 22 Sep 2020
%\cite{Schmaltz:2008vd}

\bibitem{Schmaltz:2008vd}
M.~Schmaltz and J.~Thaler,
``Collective Quartics and Dangerous Singlets in Little Higgs,''
JHEP \textbf{03} (2009), 137
doi:10.1088/1126-6708/2009/03/137
[arXiv:0812.2477 [hep-ph]].
%26 citations counted in INSPIRE as of 22 Sep 2020
%\cite{Kalyniak:2013eva}

\bibitem{Schmaltz:2010ac}
M.~Schmaltz, D.~Stolarski and J.~Thaler,
``The Bestest Little Higgs,''
JHEP \textbf{09} (2010), 018
doi:10.1007/JHEP09(2010)018
[arXiv:1006.1356 [hep-ph]].
%46 citations counted in INSPIRE as of 22 Sep 2020
%\cite{Peskin:1991sw}

\bibitem{Godfrey:2012tf}
S.~Godfrey, T.~Gregoire, P.~Kalyniak, T.~A.~W.~Martin and K.~Moats,
``Exploring the heavy quark sector of the Bestest Little Higgs model at the LHC,''
JHEP \textbf{04} (2012), 032
doi:10.1007/JHEP04(2012)032
[arXiv:1201.1951 [hep-ph]].
%11 citations counted in INSPIRE as of 22 Sep 202
%\cite{Martin:2012kqb}

\bibitem{Kalyniak:2013eva}
P.~Kalyniak, T.~Martin and K.~Moats,
``Constraining the Little Higgs model of Schmaltz, Stolarski, and Thaler with recent results from the LHC,''
Phys. Rev. D \textbf{91} (2015) no.1, 013010
doi:10.1103/PhysRevD.91.013010
[arXiv:1310.5130 [hep-ph]].
%4 citations counted in INSPIRE as of 22 Sep 2020
%\cite{Godfrey:2012tf}

\bibitem{Aranda:2018zis}
J.~I.~Aranda, D.~Espinosa-G\'omez, J.~Monta\~no, B.~Quezadas-Vivian, F.~Ram\'irez-Zavaleta, and E.~S.~Tututi, ``Flavor violation in chromo- and electromagnetic dipole moments induced by Z' gauge bosons and a brief revisit of the Standard Model", Phys.\ Rev.\ D {\bf 98}, no. 11, 116003 (2018)
  doi:10.1103/PhysRevD.98.116003
  [arXiv:1809.02817 [hep-ph]].

\bibitem{Davydychev:2000rt}
  A.~I.~Davydychev, P.~Osland and L.~Saks,
  ``Quark gluon vertex in arbitrary gauge and dimension,''
  Phys.\ Rev.\ D {\bf 63}, 014022 (2001)
  doi:10.1103/PhysRevD.63.014022
  [hep-ph/0008171].

%\cite{Aranda:2020tox}
\bibitem{Aranda:2020tox}
J.~I.~Aranda, T.~Cisneros-P\'erez, J.~Monta\~no, B.~Quezadas-Vivian, F.~Ram\'\i{}rez-Zavaleta and E.~S.~Tututi,
%``Revisiting the top quark chromomagnetic dipole moment in the SM,''
Eur. Phys. J. Plus \textbf{136} (2021) no.2, 164
doi:10.1140/epjp/s13360-021-01102-x
[arXiv:2009.05195 [hep-ph]].
%2 citations counted in INSPIRE as of 21 May 2021


\bibitem{Sirunyan:2019eyu}
A.~M.~Sirunyan \textit{et al.} [CMS],
``Measurement of the top quark forward-backward production asymmetry and the anomalous chromoelectric and chromomagnetic moments in pp collisions at $ \sqrt{s} $ = 13 TeV",
JHEP \textbf{06} (2020), 146
doi:10.1007/JHEP06(2020)146
[arXiv:1912.09540 [hep-ex]].
%3 citations counted in INSPIRE as of 22 Sep 2020

\bibitem{PDG2020}
P. A. Zyla \emph{et al.} (Particle Data Group), Prog. Theor. Exp. Phys. 2020, 083C01 (2020) and 2021 update.


%\cite{Gaitan:2015aia}
\bibitem{Gaitan:2015aia}
R.~Gaitan, E.~A.~Garces, J.~H.~M.~de Oca and R.~Martinez, ``Top quark Chromoelectric and Chromomagnetic Dipole Moments in a Two Higgs Doublet Model with CP violation,''
Phys. Rev. D \textbf{92} (2015) no.9, 094025
doi:10.1103/PhysRevD.92.094025
[arXiv:1505.04168 [hep-ph]].
%23 citations counted in INSPIRE as of 08 Jun 2021

%\cite{Martinez:2001qs}
\bibitem{Martinez:2001qs}
R.~Martinez and J.~A.~Rodriguez,
``The Anomalous chromomagnetic dipole moment of the top quark in the standard model and beyond,''
Phys. Rev. D \textbf{65} (2002), 057301
doi:10.1103/PhysRevD.65.057301
[arXiv:hep-ph/0109109 [hep-ph]].
%35 citations counted in INSPIRE as of 11 Jun 2021


\bibitem{Aboubrahim}
A. Aboubrahim, T. Ibrahim, P. Nath, and A. Zorik, Phys. Rev. D \textbf{92}, 035013 (2015).

\bibitem{Martinez:2007qf}
R.~Martinez, M.~A.~Perez and N.~Poveda,
``Chromomagnetic Dipole Moment of the Top Quark Revisited,''
Eur. Phys. J. C \textbf{53} (2008), 221-230
doi:10.1140/epjc/s10052-007-0457-6
[arXiv:hep-ph/0701098 [hep-ph]].
%42 citations counted in INSPIRE as of 22 Sep 2020
%\cite{Bermudez:2017bpx}

\bibitem{Appelquist}
T. Appelquist, M. Piai, and R. Shrock, Phys. Lett. \textbf{B 595}, 442 (2004).

%\cite{Cao:2008qd}
\bibitem{Cao:2008qd}
Q.~H.~Cao, C.~R.~Chen, F.~Larios and C.~P.~Yuan,
%``Anomalous gtt couplings in the Littlest Higgs Model with T-parity,''
Phys. Rev. D \textbf{79} (2009), 015004
doi:10.1103/PhysRevD.79.015004
[arXiv:0801.2998 [hep-ph]].
%22 citations counted in INSPIRE as of 10 Jun 2021


\bibitem{Martinez:2008hm}
R.~Martinez, M.~A.~Perez and O.~A.~Sampayo,
%``Constraints on unparticle physics from the gt $\bar{t}$ anomalous coupling,''
Int. J. Mod. Phys. A \textbf{25} (2010), 1061-1067
doi:10.1142/S0217751X10048159
[arXiv:0805.0371 [hep-ph]].


\bibitem{Ibrahim}
T. Ibrahim and P. Nath, Phys. Rev. D \textbf{84}, 015003 (2011).

\bibitem{Hayreter}
A. Hayreter and G. Valencia, Phys. Rev. D \textbf{88}, 034033 (2013).

%\cite{Hernandez-Juarez:2018uow}
\bibitem{Hernandez-Juarez:2018uow}
A.~I.~Hern\'andez-Ju\'arez, A.~Moyotl and G.~Tavares-Velasco,
%``Chromomagnetic and chromoelectric dipole moments of the top quark in the fourth-generation THDM,''
Phys. Rev. D \textbf{98} (2018) no.3, 035040
doi:10.1103/PhysRevD.98.035040
[arXiv:1805.00615 [hep-ph]].
%5 citations counted in INSPIRE as of 08 Jun 2021

%\cite{Hernandez-Juarez:2020xon}
\bibitem{Hernandez-Juarez:2020xon}
A.~I.~Hern\'andez-Ju\'arez, A.~Moyotl and G.~Tavares-Velasco,
%``Chromomagnetic and chromoelectric dipole moments of quarks in the reduced 331 model,''
[arXiv:2012.09883 [hep-ph]].
%0 citations counted in INSPIRE as of 08 Jun 2021



\bibitem{Hernandez-Juarez:2020drn}
A.~I.~Hern\'andez-Ju\'arez, A.~Moyotl, and G.~Tavares-Velasco,
``New estimate of the chromomagnetic dipole moment of quarks in the standard model,''
Eur. Phys. J. Plus \textbf{136}, 262 (2021)
doi:10.1140/epjp/s13360-021-01239-9
[arXiv:2009.11955 [hep-ph]].

%\cite{Montano-Dominguez:2021eeg}
\bibitem{Montano-Dominguez:2021eeg}
J.~Montano-Dominguez, B.~Quezadas-Vivian, F.~Ramirez-Zavaleta, E.~S.~Tututi and E.~Urquiza-Trejo, ``Off-shell chromomagnetic dipole moments in the SM at and beyond the $Z$ gauge boson mass scale",
[arXiv:2106.13923 [hep-ph]].

\bibitem{Deur:2016tte}
  A.~Deur, S.~J.~Brodsky, and G.~F.~de Teramond,
  ``The QCD Running Coupling",
  Prog.\ Part.\ Nucl.\ Phys.\  {\bf 90}, 1 (2016)
  doi:10.1016/j.ppnp.2016.04.003
  [arXiv:1604.08082 [hep-ph]].

\bibitem{Moats:2012laz}
K.~P.~Moats,
``Phenomenology of Little Higgs models at the Large Hadron Collider,''
doi:10.22215/etd/2012-09748
%0 citations counted in INSPIRE as of 22 Sep 2020
%\cite{Martinez:2007qf}

\bibitem{Martin:2012kqb}
T.~A.~W.~Martin,
``Examining extra neutral gauge bosons in non-universal models and exploring the phenomenology of the Bestest Little Higgs model at the LHC,''
doi:10.22215/etd/2012-09697
%0 citations counted in INSPIRE as of 22 Sep 2020
%\cite{Moats:2012laz}

\bibitem{Haberl:1995ek}
  P.~Haberl, O.~Nachtmann and A.~Wilch,
  ``Top production in hadron hadron collisions and anomalous top-gluon couplings,''
  Phys.\ Rev.\ D {\bf 53}, 4875 (1996)
  doi:10.1103/PhysRevD.53.4875
  [hep-ph/9505409].

\bibitem{Khachatryan:2016xws}
  V.~Khachatryan {\it et al.} [CMS Collaboration],
  ``Measurements of t t-bar spin correlations and top quark polarization using dilepton final states in pp collisions at sqrt(s) = 8 TeV,''
  Phys.\ Rev.\ D {\bf 93}, no. 5, 052007 (2016)
  doi:10.1103/PhysRevD.93.052007
  [arXiv:1601.01107 [hep-ex]].

\bibitem{Bernreuther:2013aga}
  W.~Bernreuther and Z.~G.~Si,
  ``Top quark spin correlations and polarization at the LHC: standard model predictions and effects of anomalous top chromo moments,''
  Phys.\ Lett.\ B {\bf 725}, 115 (2013)
  Erratum: [Phys.\ Lett.\ B {\bf 744}, 413 (2015)]
  doi:10.1016/j.physletb.2013.06.051, 10.1016/j.physletb.2015.03.035
  [arXiv:1305.2066 [hep-ph]].

\bibitem{Choudhury:2014lna}
  I.~D.~Choudhury and A.~Lahiri,
  ``Anomalous chromomagnetic moment of quarks,''
  Mod.\ Phys.\ Lett.\ A {\bf 30}, no. 23, 1550113 (2015)
  doi:10.1142/S0217732315501138
  [arXiv:1409.0073 [hep-ph]].

\bibitem{Czarnecki:1997bu}
  A.~Czarnecki and B.~Krause,
  ``Neutron electric dipole moment in the standard model: Valence quark contributions,''
  Phys.\ Rev.\ Lett.\  {\bf 78}, 4339 (1997)
  doi:10.1103/PhysRevLett.78.4339
  [hep-ph/9704355].


\bibitem{SNSB}
G. Aad \emph{et. al} [ATLAS Collaboration], ``Search for a heavy Higgs boson decaying into a Z boson and another heavy Higgs boson in the $\ell \ell bb$ and $\ell \ell WW$ final states in $pp$ collisions at $\sqrt{s}=13$ $\text {TeV}$ with the ATLAS detector", Eur. Phys. J. C 81, 396 (2021).



\bibitem{CMS:2018hff}
A.~M.~Sirunyan \textit{et al.} (CMS),
``Search for high-mass resonances in final states with a lepton and missing transverse momentum at $ \sqrt{s}=13 $ TeV,''
JHEP \textbf{06}, 128 (2018)
doi:10.1007/JHEP06(2018)128
[arXiv:1803.11133 [hep-ex]].
%

\bibitem{ATLAS:2018ihk}
M.~Aaboud \textit{et al.} (ATLAS),
``Search for High-Mass Resonances Decaying to $\tau\nu$ in pp Collisions at $\sqrt{s}$=13  TeV with the ATLAS Detector,''
Phys. Rev. Lett. \textbf{120}, no.16, 161802 (2018)
doi:10.1103/PhysRevLett.120.161802
[arXiv:1801.06992 [hep-ex]].

\bibitem{ATLAS:2019lsy}
G.~Aad \textit{et al.} (ATLAS),
``Search for a heavy charged boson in events with a charged lepton and missing transverse momentum from $pp$ collisions at $\sqrt{s} = 13$ TeV with the ATLAS detector,''
Phys. Rev. D \textbf{100}, no.5, 052013 (2019)
doi:10.1103/PhysRevD.100.052013
[arXiv:1906.05609 [hep-ex]].
%

\bibitem{CMS:2018fza}
A.~M.~Sirunyan \textit{et al.} (CMS),
``Search for a W'  boson decaying to a $\tau$ lepton and a neutrino in proton-proton collisions at $\sqrt{s} =$ 13 TeV,''
Phys. Lett. B \textbf{792}, 107-131 (2019)
doi:10.1016/j.physletb.2019.01.069
[arXiv:1807.11421 [hep-ex]].
%
\end{thebibliography}
\end{document}